\documentclass[lettersize,journal]{IEEEtran}
\usepackage{graphicx}
\usepackage{xcolor}
\usepackage{todonotes}

\usepackage[hyphens]{url}
\usepackage{hyperref}
\usepackage[hyphenbreaks]{breakurl}
\usepackage{amsmath}
\usepackage{amsthm}
\usepackage{booktabs}
\usepackage{multirow}
\usepackage{subfigure}
\usepackage{amssymb}
\usepackage[mathlines]{lineno}
\usepackage{subfigure}
\usepackage[linesnumbered, ruled]{algorithm2e}
\usepackage{lipsum}

\makeatletter
\newcommand{\nosemic}{\renewcommand{\@endalgocfline}{\relax}}
\newcommand{\dosemic}{\renewcommand{\@endalgocfline}{\algocf@endline}}
\let\oldnl\nl
\newcommand{\nonl}{\renewcommand{\nl}{\let\nl\oldnl}}
\makeatother

\AtBeginDocument{%
  \providecommand\BibTeX{{%
    \normalfont B\kern-0.5em{\scshape i\kern-0.25em b}\kern-0.8em\TeX}}}

\begin{document}

\title{TinyML Design Contest for Life-Threatening Ventricular Arrhythmia Detection}

\author{Zhenge~Jia,~\IEEEmembership{Member,~IEEE,}
        Dawei~Li,~\IEEEmembership{Member,~IEEE,}
        Cong~Liu,
        Liqi~Liao,
        Xiaowei~Xu,~\IEEEmembership{Member,~IEEE,}
        Lichuan~Ping,
        ~and~Yiyu~Shi,~\IEEEmembership{Senior~Member,~IEEE}
\thanks{Z. Jia and Y. Shi are with the Department
of Computer Science and Engineering, University of Notre Dame, Notre Dame, IN, 46556 USA 
(e-mail: \{zjia2, yshi4\}@nd.edu).}
\thanks{D. Li, C. Liu, and L. Liao are with South Central Minzu University, Wuhan, 430074, China (e-mail: leedavidhust@outlook.com, liucong16948@foxmail.com, liaoliqi2022@163.com).}
\thanks{X. Xu is with the Department of Cardiovascular Surgery, Guangdong Cardiovascular Institute, Guangdong Provincial People's Hospital (Guangdong Academy of Medical Sciences), Southern Medical University, Guangzhou, Guangdong 510080, China (e-mail: xiao.wei.xu@foxmail.com).}
\thanks{L. Ping is with Singular Medical Co., Ltd, Suzhou, Jiangsu, 215163, China (e-mail: pinglichuan@singularmedical.net).}
\thanks{Please address comments to leedavidhust@outlook.com and yshi4@nd.edu.}
}

\markboth{IEEE Transactions on Computer Aided Design of Integrated Circuits \& Systems,~Vol.~1, No.~1, April~2023}%
{Jia \MakeLowercase{\textit{et al.}}: A Sample Article Using IEEEtran.cls for IEEE Journals}

\maketitle

\begin{abstract}
The first ACM/IEEE TinyML Design Contest (TDC) held at the 41st International Conference on Computer-Aided Design (ICCAD) in 2022 is a challenging, multi-month, research and development competition. TDC'22 focuses on real-world medical problems that require the innovation and implementation of artificial intelligence/machine learning (AI/ML) algorithms on implantable devices. The challenge problem of TDC'22 is to develop a novel AI/ML-based real-time detection algorithm for life-threatening ventricular arrhythmia over low-power microcontrollers utilized in Implantable Cardioverter-Defibrillators (ICDs). The dataset contains more than 38,000 5-second intracardiac electrograms (IEGMs) segments over 8 different types of rhythm from 90 subjects. The dedicated hardware platform is NUCLEO-L432KC manufactured by STMicroelectronics. TDC'22, which is open to multi-person teams world-wide, attracted more than 150 teams from over 50 organizations. This paper first presents the medical problem, dataset, and evaluation procedure in detail. It further demonstrates and discusses the designs developed by the leading teams as well as representative results. This paper concludes with the direction of improvement for the future TinyML design for health monitoring applications. 
\end{abstract}


\section{Introduction}

The 41st International Conference on Computer-Aided Design (ICCAD) held its first TinyML Design Contest (TDC) in 2022. 
TDC'22 motivates participating teams to design artificial intelligence and machine learning (AI/ML) algorithms for life-threatening ventricular arrhythmia detection over intracardiac electrograms (IEGMs), and deploy the algorithm on a low-power and resource-constrained microcontroller in Implantable Cardioverter-Defibrillators (ICDs). 
TDC'22 provides the opportunity to explore various TinyML designs and compare the innovative designs in terms of comprehensive performances over unified metrics and platforms.

The aim of TDC'22 is to address real industry needs and further save more lives through AI/ML. 
Ventricular fibrillation (VF) and ventricular tachycardia (VT) are both critical life-threatening ventricular arrhythmias (VAs), which are the main cause of Sudden Cardiac Death (SCD)~\cite{santini2007primary, adabag2010sudden}. 
People at high risk of SCD rely on ICDs to deliver proper and in-time defibrillation treatment when experiencing life-threatening VAs~\cite{santini2007primary, adabag2010sudden}. 
However, existing industry practice is simple-rule based detection, and the detection methods have not changed over the past few decades~\cite{zanker2016tachycardia, madhavan2013optimal}.  
AI/ML based detection methods can potentially revolutionize ICD design by extracting features not easily identifiable by or even unknown to human experts while reducing the required expertise in methodology design.

TDC'22 is the first TinyML contest in healthcare. 
The participants are asked to design and implement a working and open-source AI/ML algorithm that can automatically discriminate life-threatening VAs (i.e., binary classification: VAs or non-VAs) from other heart rhythms. 
Moreover, different from the existing biosignal challenges which focus only on detection accuracy, the proposed AI/ML algorithm should be able to be deployed and executed on the dedicated microcontroller (MCU) NUCLEO-L432KC development kit of the STM32 series~\cite{STM32L432KC}. 
The submitted design is evaluated with comprehensive performance metrics in terms of detection accuracy, inference latency, and flash memory usage. 
Therefore, TDC'22 focuses on developing an AI/ML algorithm that achieves high accuracy, low inference latency, and low memory footprint for life-threatening ventricular arrhythmia detection on ICDs.
TDC'22 takes into full consideration the requirements of ICDs: Accurate and real-time data processing on low-power and extremely resource-constrained embedded devices.

TDC'22 provides a large IEGM dataset as well as great flexibility in AI/ML algorithm development such that the participating teams could design and deploy algorithms on board with little hardware programming experience and medical knowledge. 
The whole dataset contains more than 38,000 IEGM segments with manual labels provided by SingularMedical.
The publicly released dataset includes over 30,000 IEGMs segments from 76 subjects while the testing dataset remains private for evaluation purposes. 
TDC'22 further developed a unified evaluation method to evaluate the submitted design in terms of comprehensive metrics. 
With the training dataset and unified evaluation method, participating teams could either train their models/algorithms using prevalent deep learning frameworks (e.g., TensorFlow, PyTorch) and deploy the saved models/algorithms on MCU board with the C code auto-generated by STM32CubeMX~\cite{CubeAI}. 
Or they could develop and deploy the AI/ML algorithms on board using their own framework or purely C code. 
Participating teams could even try a hybrid development method, that is, develop their algorithms with prevalent deep learning frameworks and modify the auto-generated C code by STM32CubeMX to further optimize the practical performances. 
With the great flexibility in algorithm design and deployment, the winning designs are very innovative (beyond deep neural networks) and could achieve a comparable or even beat the current industry standards.

This paper describes TDC'22 in terms of various perspectives. 
It first introduces the settings of the contest in detail, including the contest problem with the medical background, dataset, and evaluation. 
Furthermore, it presents the results of the winning teams, a comprehensive analysis of the innovative design, and a summary of improvements for the future TinyML design contest. 
The insight and lessons are discussed for the future development of TinyML design in healthcare, especially for health monitoring applications. 
Particularly, we will elaborate Artificial-Human Intelligence co-design for efficient processing and accurate detection on embedded medical devices. 

The training dataset, the source code of top-8 teams together with the unified evaluation method, and additional information about the contest can be accessed at \url{https://tinymlcontest.github.io/TinyML-Design-Contest/index.html}.

The rest of the paper is organized as follows:  
Section~\ref{sec_related} introduces the related works about previous contests and the corresponding benchmarking datasets. 
Section~\ref{sec_contest} describes the details of the contest problem and the related medical background. 
Section~\ref{sec_data} presents the dataset in terms of various characteristics. 
Section~\ref{sec_evaluation} demonstrates the comprehensive performance metrics, hardware platform, and evaluation method. 
The analysis of the teams' designs is presented in  Section~\ref{sec_design}. 
Section~\ref{sec_results} details the results of the teams. 
Finally, we conclude the paper with a discussion of the contest and future improvement for TinyML in healthcare in Section~\ref{sec_conclusion}.

\section{Related Works}
\label{sec_related}

In this section, we will introduce the related works about previous contests focusing on biosignal data and the contests focusing on TinyML. 
The datasets provided by these contests will also be introduced. 

\subsection{Contests in Health Monitoring}
While a majority of the AI/ML contests in healthcare focus on medical images~\cite{dsb17, PAIP, luna16, isbi-info, brats}, biosignal plays an important role in health condition monitoring. 
Different from medical images which require  extensive acquisition skills, biosignal data is relatively easy to be obtained via wearable or implantable devices.
Nevertheless, the volume of biosignal data can be enormously large since the data mostly come from continuous monitoring. 
Accurately labeling these data is an overwhelming task for physicians even with the help of computer-aided methods. 
In addition, instead of revealing the medical condition through a computer-vision perspective, the algorithms designed for health monitoring could only rely on the information far from rich (i.e., the signal data of a few channels). 
These factors make the contests in health monitoring unique and challenging from the others.

There are several contests contributing to health monitoring over various types of biosignals. 
The George B. Moody PhysioNet Challenge, which was first launched in 2000, is an annually held contest to tackle clinically interesting questions that are either highly topical or neglected~\cite{physionet, goldberger2000physiobank}. 
Biosignal has been utilized as the challenge data for a majority of recent PhysioNet challenges. 
The PhysioNet Challenges 2016~\cite{clifford2016classification, liu2016open} and 2022~\cite{reyna2022heart, oliveira2021circor} both utilize heart sound signal, which is known as Phonocardiogram (PCG), 
In 2016, the challenge focused on classifying normal versus abnormal heart sound episodes from a single short recording from a single precordial location~\cite{physionet16, clifford2016classification}. 
The dataset consisting of five databases contains a total of 3,126 heart sound recordings, which last from 5 to 120 seconds. 
In 2022, the challenge focused on detecting the presence or absence of murmurs from multiple heart sound recordings from multiple auscultation locations~\cite{physionet22, reyna2022heart}. 
The dataset provided by The PhysioNet Challenges 2022 consists of 5,272 heart sound recordings from 1,568 pediatric patients collected in rural Brazil~\cite{physionet22, oliveira2021circor}. 
The PhysioNet Challenges 2021~\cite{physionet21, reyna2021will} focused on classifying cardiac abnormalities from 12-lead Electrocardiogram (ECG), which is another essential type of biosignal non-invasively reflecting the electrical activity of the heart~\cite{physionet21}.
The submitted algorithm is required to identify a set of one or more classes with the corresponding probability score.
The dataset provided by the challenge consists of 7 databases, which include over 100,000 12-lead ECG recordings~\cite{physionet21, reyna2021will}. 

Apart from The George B. Moody PhysioNet Challenge, there are also some challenges organized for health monitoring applications over biosignals. 
The China Physiological Signal Challenge (CPSC), which was first launched in 2018, is a competition focusing on the automatic identification of the rhythm/morphology abnormalities in 12-lead ECGs~\cite{CSPC18, liu2018open}. 
The provided dataset was collected from 11 hospitals and it contains 9,831 12-lead ECG recordings lasting from 6 to 60 seconds. 
In addition, Challenge UP 2019 was organized for fall detection based on a public multimodal dataset~\cite{up19}. 
The dataset was collected from 12 subjects that performed 11 activities and falls using inertial measurement unit (IMU) sensors and vision devices.

\subsection{Contests in TinyML}

A medical device is intended for use in either the diagnosis of disease or in the cure, mitigation, treatment, and prevention of disease. Typical medical devices include pacemakers, the heart-lung machine, artificial organs, cochlear implants, ocular prosthetics, facial prosthetics, somato prosthetics, etc. Microcontroller and Application Specific Integrated Circuits (ASICs) are often applied to these devices for computing purposes and energy efficiency.

TinyML is a cutting-edge field that brings AI/ML to the resource-constrained domain of tiny devices and embedded systems~\cite{tinyml}.
The contests in TinyML mainly focus on computer vision. 
The tinyML Foundation launched a contest "Eyes on Edge: tinyML Vision Challenge" to build advanced applications with low-power machine learning inferencing and computer vision for edge devices~\cite{eyesonedge}. 
There are no specific requirements for the application, but to devise a proof of concept using machine vision on embedded systems to address an industry-grade problem~\cite{eyesonedge}.
The hardware platforms suggested include OpenMV Cam H7, Arduino Portenta Vision Shield, Raspberry Pi, etc.  
Another TinyML Challenge 2022 held by AI4G is to create a low-cost, low-power, reliable, accurate weather station~\cite{weather}.
The goal of the contest is to utilize the TinyML technique to measure all weather conditions with a focus on rain and wind and deploy it on board.

In summary, the existing contests on health monitoring applications focus solely on the accurate and automatic detection of the event with non-invasive monitors. 
The computational platform in these contests is generally assumed to be a server with adequate resources to adopt the submitted algorithms. 
As for implantable devices such as ICDs, real-time detection is critical, and the detection for the rapid alert is expected to be conducted on the embedded monitoring device. 
The resource constraint of the embedded implantable device is another key factor to be considered in the algorithm design in addition to accuracy. 
Before the launch of the TinyML contest, there is still no contest in the healthcare field targeting comprehensive detection performances (i.e., real-time and accurate) for invasive health monitoring. 
\section{Contest Problem}
\label{sec_contest}

In this section, the medical background of the contest is first introduced, followed by the introduction of the contest objective.

\subsection{Contest Background}

Heart disease is the first leading cause of death in the US, according to the 2021 CDC Mortality Report~\cite{ahmad2022provisional}.
VF and VT are two types of life-threatening ventricular arrhythmias (VAs), which are the main cause of SCD~\cite{santini2007primary, adabag2010sudden}. 
More than 60\% of deaths from cardiovascular disease are from out-of-hospital SCD~\cite{adabag2010sudden}. 
It is caused by a malfunction in the heart's electrical system that can occur when the lower chambers of the heart suddenly start beating in an uncoordinated fashion, preventing the heart from pumping blood out to the lungs and body~\cite{adabag2010sudden}. 
Unless the heart is shocked back into normal rhythm, the patient rarely survives. 

People at high risk of SCD rely on ICDs instead of Automated External Defibrillator (AED) to deliver proper and in-time defibrillation treatment when experiencing life-threatening VAs. 
ICD is a small device implanted typically under the skin in the left upper chest and is programmed to release defibrillation (i.e., electrical shock) therapy on VT and VF to restore rhythm back to normal. 
The early and in-time diagnosis and treatment of life-threatening VAs can increase the chances of survival.

However, current detection methods on ICDs are just simple rule-based detection, which has not been updated for several decades. 
The VA detection approach is based on a wide variety of criteria obtained from clinical trials, and there are hundreds of programmable criteria parameters affecting the defibrillation delivery decision~\cite{zanker2016tachycardia, madhavan2013optimal}.
To obtain optimal VA detection performances, physicians must make much effort to adjust those parameters based on their experiences through frequent follow-ups. 
This parameter adjustment process also causes a high economic burden for patients.

Automatic detection with less expertise involved in ICDs can further improve detection performances and reduce the workload from physicians in criteria design and parameter adjustment~\cite{acharya2018automated, jia2021ijcai}. 
Detection methods based on artificial intelligence and machine learning (AI/ML) have demonstrated the potential to revolutionize ICD detection. 
The invocation of AI/ML can potentially revolutionize ICD design by extracting features not easily identifiable by or even unknown to human experts. 
The design process also reduces the dependency on domain expertise.
In this case, AI/ML-based VA detection on ICDs could significantly improve the clinical outcomes of ICD recipients from both treatment and economy perspectives.

\subsection{Contest Objective}
The goal of TCD'22 is to design an AI/ML algorithm that can identify life-threatening VAs from single-lead (i.e., RVA-Bi) IEGM recordings. 
The participants are asked to design and implement a working and open-source AI/ML algorithm that, based on the IEGM recordings sensed by the single-chamber ICDs, can automatically and accurately identify life-threatening VAs while satisfying the requirements of in-time detection on the resource-constrained microcontroller (MCU) utilized in ICDs.  
The effectiveness of the AI/ML algorithm is evaluated on the practical MCU board by comprehensive metrics including accuracy, latency, and memory footprint. 
More details about the dataset and evaluation will be presented in Section~\ref{sec_data} and Section~\ref{sec_evaluation}. 

\section{Dataset}
\label{sec_data}

\begin{table}[t]
\renewcommand\arraystretch{1.3}
\caption{Label and Corresponding Rhythm Type}
\centering
\setlength{\tabcolsep}{4mm}{
\begin{tabular}{l|l}
\hline
\textbf{Label} & \textbf{Rhythm Type}                  \\ \hline
AFb            & Atrial Fibrillation                   \\
AFt            & Atrial Flutter                        \\
SR             & Sinus Rhythm                          \\
SVT            & Supraventricular Tachycardia          \\
VFb            & Ventricular Fibrillation              \\
VFt            & Ventricular Flutter                   \\
VPD            & Ventricular Premature Depolarization \\
VT             & Ventricular Tachycardia               \\ \hline
\end{tabular}
}
\label{table-label}
\end{table}

\begin{figure*}[t]
\includegraphics[width = 0.92\textwidth]{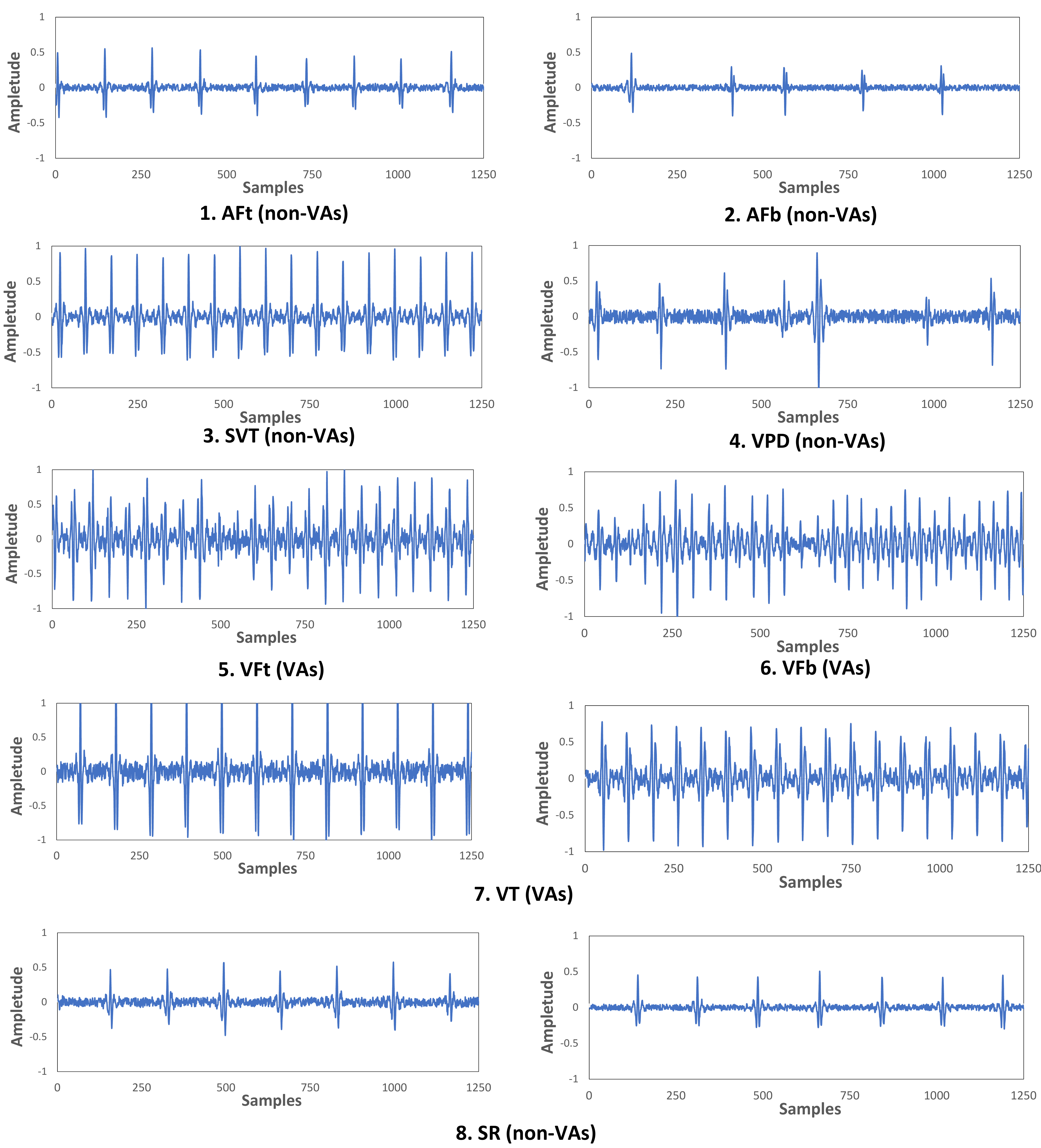}
\centering
\caption{Overview of the dataset. There are two main categories: Life-threatening VAs and Non-VAs. Each category includes several sub-categories as shown in the bracket of the sub-caption. 
}
\label{fig-segs}
\end{figure*}

The dataset is constructed based on the IEGM recordings provided by SingularMedical~\cite{singular}. 
The IEGM recordings are retrieved from the lead RVA-Bi of a single-chamber ICD from 90 subjects. 
The sampling rate of recordings is 250 Hz. 
Each recording is first partitioned into episodes corresponding to the rhythm label provided by cardiologists. 
As a result, each episode is with only one label that describes the cardiac abnormalities (or normal sinus rhythm). 
Next, each episode is segmented by a 5-second sliding window with a stride of 150 sampling points. 
Each segment is with a fixed length of 1250 sampling points.

Segments are with the labels of 2 main categories (i.e., VAs and Non-VAs) and 8 sub-categories. 
The full list of labels and the specific rhythm types is shown in Table~\ref{table-label}.
For the category of life-threatening VAs, the corresponding sub-category labels include VT (Ventricular Tachycardia), VFb (Ventricular Fibrillation), and VFt (Ventricular Flutter, a special type of tachycardia affecting the ventricles). 
For the non-VAs, the segments are labeled with the sub-category label other than VT, VFb, VFt as shown in Table~\ref{table-label}. 
Fig.~\ref{fig-segs} demonstrates the morphological characteristics of the IEGM segments with the caption indicating the main category in bracket and the sub-category right after the number.  
As shown in the Figure, the most prominent characteristic of the segments with the label in VAs category is the fast heart rate, which is represented by the number of peaks. 
While the segments with the label in non-VAs category demonstrate a much lower heart rate, the number of peaks in SVT segment is still comparable to the VAs segments. 
Such characteristics make it even harder to accurately discriminate SVT from VAs and may lead to inappropriate shock in the ICD scenarios.

\begin{figure}[t]
\includegraphics[width = 0.42\textwidth]{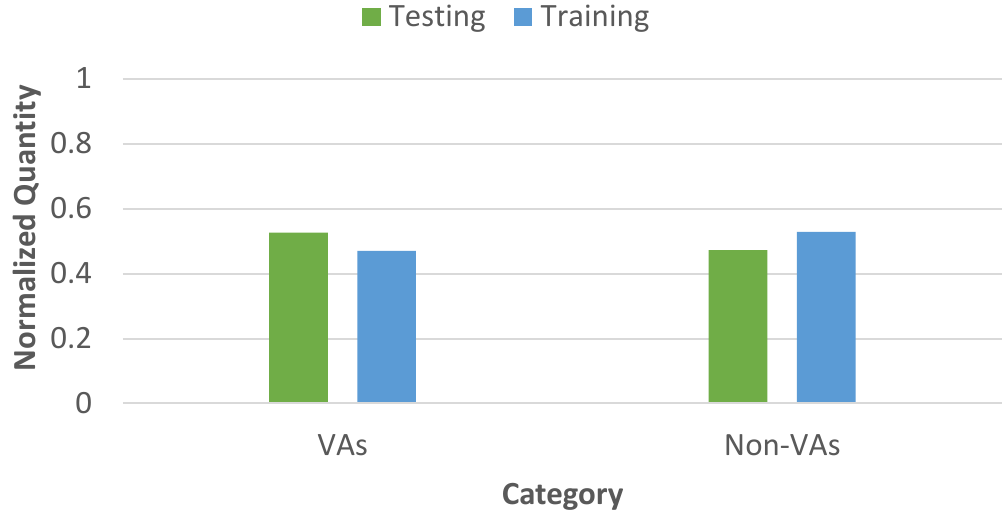}
\centering
\caption{Distribution of the training and testing dataset with respect to main categories. 
}
\vspace{-10pt}
\label{fig-dis}
\end{figure}

\begin{figure}[t]
\centering
\subfigure[VAs.]{
\includegraphics[width =0.21\textwidth]{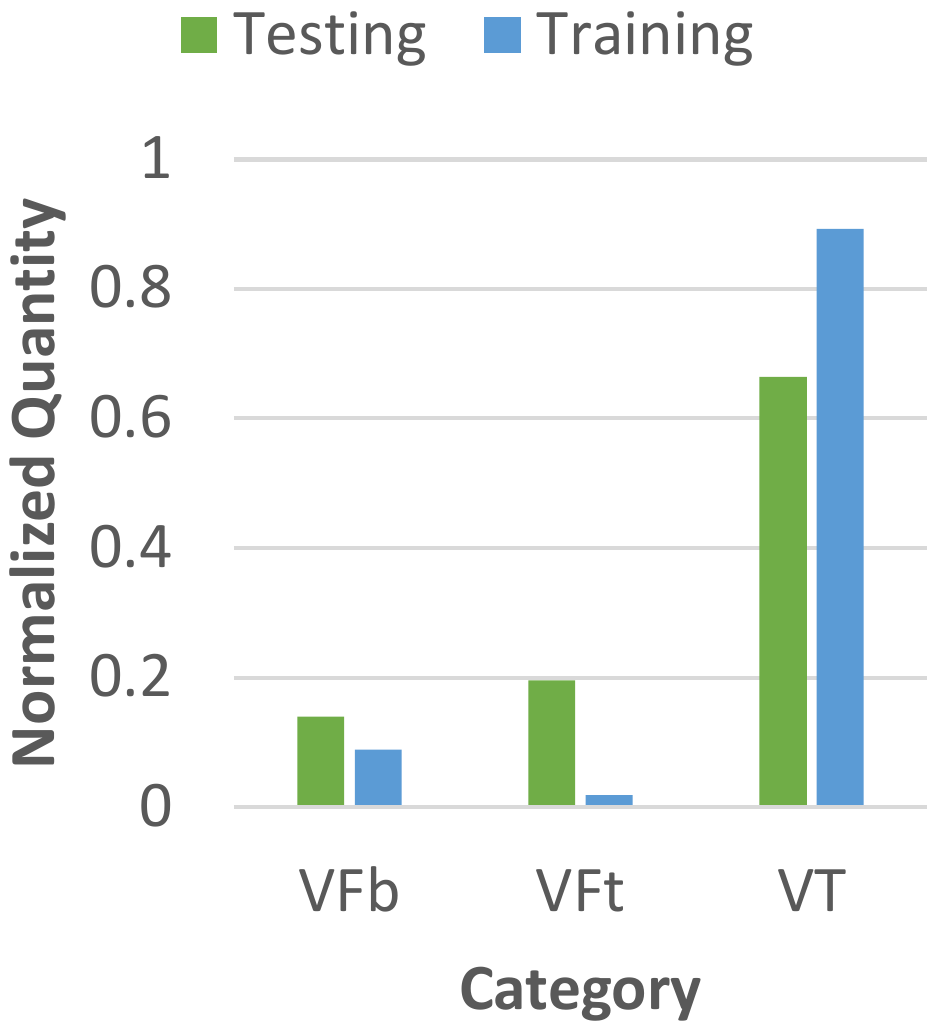} 
\label{fig-dis-va} \vspace{-5pt}}
\subfigure[Non-VAs.]{
\includegraphics[width =0.21\textwidth]{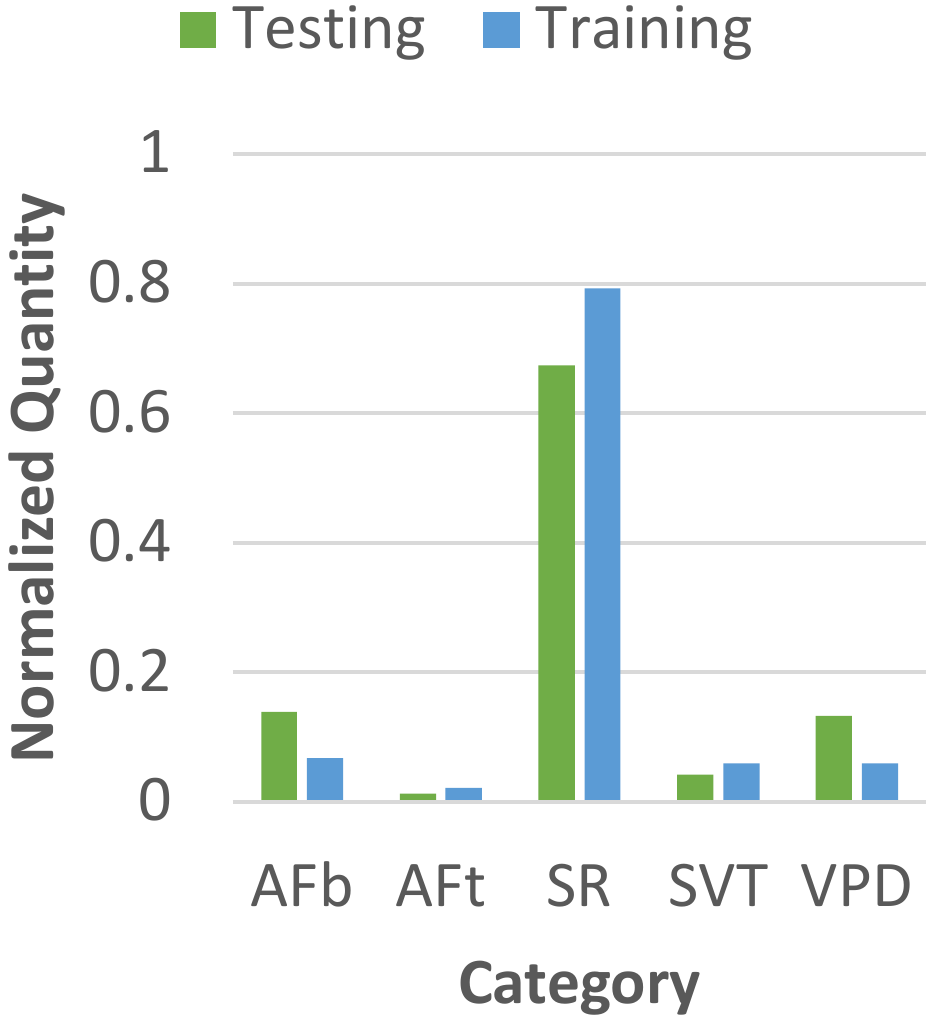}
\label{fig-dis-nonva}}
\caption{Distribution of the training and testing dataset with respect to sub-categories of VAs and non-VAs.}
\label{fig-dis-va-nonva}
\vspace{-10pt}
\end{figure}

The IEGM segments are partitioned patient-wisely in the training and testing set. 
The aim of the patient-wise partition is to ensure that the subject's data is not mixed in these two sets. 
In other words, once the patient is selected for training or testing, all the corresponding segments would be utilized only for one purpose.
In the contest, 85\% of the subjects’ IEGM segments are released as training material. 
The rest 15\% of the subjects’ IEGM segments will be utilized to evaluate the detection performances of the submitted AI/ML algorithms. 
The dataset for final evaluation will remain private and will not be released.

The distribution of the training and testing dataset with respect to the main category is shown in Fig.~\ref{fig-dis}, where the Y-axis represents the ratio of the quantity of two groups of IEGM segments with the same label from the training and testing partition.  
As shown in the figure, the patient-wise partition also maintains a relatively balanced distribution of each category in the testing and training dataset.
In addition, Fig.~\ref{fig-dis-va-nonva} demonstrates the distribution of the training and testing dataset with respect to each sub-category of VAs and Non-VAs. 
The Y-axis of Fig.~\ref{fig-dis-va-nonva} again represents the ratio of the quantity of two groups of IEGM segments with the same label from the training and testing partition.
Though there is imbalanced distribution in the sub-category such as VFt, Fig.~\ref{fig-dis-va-nonva} again indicates an overall balanced distribution of testing and training datasets. 
The aim of applying patient-wise partition is to motivate the teams to propose AI/ML algorithms that work for practical scenarios where the patient health data is non-independent and identically distributed (non-iid).

\section{Evaluation}
\label{sec_evaluation}

In this section, the designated hardware platform and design setup are first introduced. 
The evaluation metrics, which are utilized to measure the comprehensive performances of each team's design, will be introduced. 
The evaluation method, which provides teams with great flexibility in method design based on the practical hardware platform, will finally be presented. 

\subsection{Hardware Platform and Design Setup}

The hardware platform, development kit NUCLEO-L432KC~\cite{STM32L432KC}, was designated to the participating teams in TDC'22. 
Particularly, the \$10.99 development board is equipped with an ARM Cortex-M4 core at 80 MHz, 256 KiB of flash memory, 64 KiB of SRAM, and an embedded ST-LINK/V2-1 debugger/programmer. 
The power consumption of the board is around 30 mW in operation and 1.5 mW in idle. 
The hardware platform targets ultra-low-power embedded computation and is suitable for computational tasks in ICDs. 
The development board also supports STM32 X-Cube-AI, which is an STM32Cube Expansion Package part of the STM32Cube~\cite{CubeAI}. 
STM32 X-Cube-AI is part of the STM32 AI ecosystem and extends STM32CubeMX capabilities with automatic conversion of pre-trained Artificial Intelligence algorithms, including Neural Network and classical Machine Learning models, and integration of the optimized library into the user's project~\cite{CubeAI}. 

One major advantage of NUCLEO-L432KC is to provide the team with little hardware programming experience to participate in the contest to deploy their TinyML design on board.
Teams could train and save their neural network model with any preferred framework on the server, and deploy the model on the MCU board with the help of X-Cube-AI embedded in STM32CubeMX. 
In TDC'22, we provide participating teams with a step-by-step guideline on the model deployment available at \url{https://github.com/tinymlcontest/tinyml_contest2022_demo_example/blob/master/README-Cube.md}. 
This guideline also pushes forward the understanding and practical implementation of TinyML design.

\subsection{Evaluation Metrics}
The metrics are based on detection accuracy, inference latency, and memory footprint to comprehensively evaluate the effectiveness and practicability of the submitted design. 
Power consumption is not included in the metrics since the designated MCU board runs with relatively stable power consumption in operation and 
the lower inference latency indicates a lower energy consumption.

For detection accuracy, $F_{\beta}$ score is utilized as the metric. 
The idea is to utilize a single metric that weights the two ratios (precision and recall) in a balanced way, requiring both to have a higher value for $F_{\beta}$ score. 
To be more specific, the confusion matrix is first computed over the classification results of life-threatening VAs and non-VAs returned by the AI/ML algorithm over the IEGM segments of the testing dataset.
True Positive (TP), True Negative (TN), False Positive (FP), and False Negative (FN) can be then derived from the confusion matrix. 
Next, recall is calculated as follows:
\begin{equation} \label{func-recall}
Recall = \frac{TP}{TP+FN}. 
\end{equation}
Precision is calculated as follows:
\begin{equation} \label{func-precision}
Precision = \frac{TP}{TP+FP}. 
\end{equation}
The $F_{\beta}$ score can be calculated based on recall and precision as follows:
\begin{equation} \label{func-fb}
F_{\beta} = (1+{\beta}^2) \times \frac{Precision \cdot Recall}{({\beta}^2 \cdot Precision) + Recall}, 
\end{equation}
where $\beta = 2$. 
This value gives a higher weight to recall since the detection accuracy of life-threatening VAs is the most critical function for ICDs. 
The setting of $F_{\beta}$ score is expected to discriminate as many VAs as possible to avoid missing shock on VAs while trying to boost the detection accuracy on non-VAs to reduce inappropriate shock rate (i.e., the shock triggered by the episodes which are incorrectly classified as VAs). 
The best value of $F_{\beta}$ score is at 1 and the worst value is at 0.

The metric for latency is based on the average latency $L$ (in $ms$) of inferences executed on the designated board.
The latency score $L_n$ is calculated as follows:
\begin{equation} \label{func-latency}
L_n = 1-\frac{L-L_{min}}{L_{max}-L_{min}}, 
\end{equation}
where $L_{min} = 1$ and $L_{max} = 200$ as the normalized bound. 
For real-time VA detection on ICDs, the maximum latency requirement in this contest is set to 200 $ms$. 
The minimum latency execution is set as 1 $ms$ based on the estimate of the computational capability of the MCU board. 
We further set the rule that the highest $L_n$ score achieved by the team is 1 and the lowest $L_n$ score is 0.

The metric of memory footprint is based on the flash occupation $M$ (in $KiB$) of the program on the board. 
The memory footprint score $M_n$ is calculated as follows:
\begin{equation} \label{func-memory}
M_n = 1-\frac{M-M_{min}}{M_{max}-M_{min}}, 
\end{equation}
where $M_{min} = 5$ and $M_{max} = 256$ as the normalized bound. 
The maximum memory occupation is set to 256 KiB since the memory capacity of the development board is 256 KiB. 
The minimum memory occupation is set to 5 KiB since the initial program code occupied around 5 KiB. 
We further set the rule that the highest $M_n$ score achieved by the team is 1 and the lowest $M_n$ score is 0.

The final score $FS$ is a weighted combination of detection accuracy, inference latency, and memory footprint as follows:
\begin{equation} \label{func-final-score}
FS = 100 \cdot F_{\beta} + 20 \cdot L_{n} + 20 \cdot M_{n},
\end{equation}
where the total score of $FS$ is 140. 
The higher value of $FS$ represents a better comprehensive performance. 
The final score puts more weight on the detection accuracy since the precise discrimination of life-threatening VAs is the most critical function of an ICD. 
The final score could also benefit from the design by lowering the inference latency and memory footprint of the AI/ML algorithm design. 
We set an equal importance on memory footprint as latency because the memory capacity for storing health data records is also critical. Especially for implantable devices, due to the extremely constrained physical size, the memory capacity spared for health data recording is always limited. On the other hand, it is always demanding to acquire more health data for model parameters tailored for the patient for better detection performances. Therefore, the less memory footprint occupied by the AI/ML model and corresponding program, the more health data can be stored.  
Furthermore, the marginal effects on the final score have been applied to the metrics in inference latency and memory footprint by the normalization in Eqn.~\ref{func-latency} and Eqn.~\ref{func-memory}. 
In this way, the optimization focus would be directed toward the practicability (i.e., precise VA detection while fitting the resource-constraint device) of the AI/ML algorithm on ICDs. 
The details of the setting's benefits would be illustrated with example submission and results in Section~\ref{sec_results}.

In healthcare, it is critical to bring the best detection performance for each individual. 
The metric reflecting the generalization of the trained AI/ML model should be considered in the future. 
Due to individual differences, the morphological characteristics of IEGMs (with the same label) are inter-patiently variable. 
This can lead to an uneven detection performance over patients using the trained AI/ML model. 
Therefore, a metric that can measure the generalization of the trained AI/ML model across the patients is demanded.

\begin{figure}[t]
\includegraphics[width = 0.47\textwidth]{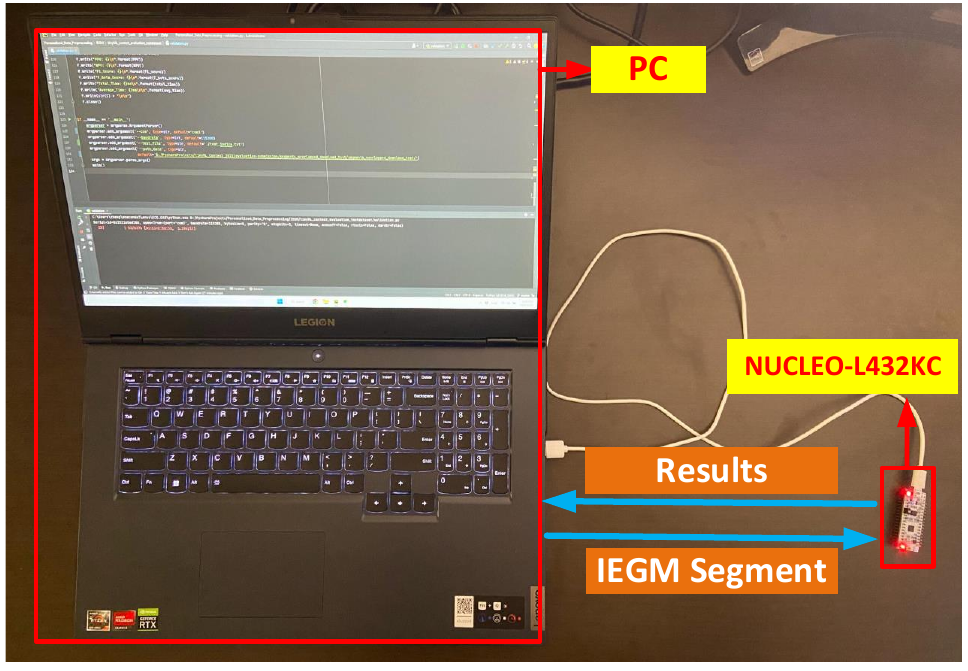}
\centering
\caption{The evaluation platform consists of a PC and a NUCLEO-L432KC board.
}
\vspace{-10pt}
\label{fig-platform}
\end{figure}

\subsection{Evaluation Method}

The final evaluation is mainly conducted on the designated MCU board to obtain all required metrics.
In addition to the accurate measurement, participating teams are provided with great flexibility for their algorithm design and deployment. 
In other words, participating teams are free to choose any framework to design the AI/ML algorithms, and any package to deploy the model on board instead of being restricted to X-Cube-AI.
Teams are also unconstrained to apply any approach to optimize the model to improve the comprehensive performances depending on their level of expertise in hardware programming. 
In this way, participating teams, no matter how much hardware programming experience gained before the contest, can freely explore the innovation in the design without the interference of the model deployment.

To conduct a fair and universal performance comparison, we set up an evaluation platform and provide a code framework in Keil to uniformly evaluate the submitted AI/ML algorithm. 
The evaluation platform is based on a NUCLEO-L432KC board and a PC as shown in Fig.~\ref{fig-platform}. 
What participating teams should do is to implement an interface function \textit{aiRun} to execute the model inference in the provided code framework in Keil. 
During the evaluation, the PC sends each IEGM segment to the board and the board sends the measured metrics back to the PC. 
The communication between the board and PC is based on UART. 
The provided code framework and guideline can be accessed at \url{https://github.com/tinymlcontest/tinyml_contest2022_demo_evaluation}. 

For accuracy, the board conducts the inference on the received IEGM segment and sends the inference result back to the PC. 
On the PC side, the confusion matrix is established and $F_{\beta}$ can be obtained until all segments are iteratively sent to the board. 
For inference latency, the SysTick timer on board is utilized to record the starting and ending time ticks of the interface function for inference. 
The time period is transmitted back to the PC and accumulated as the total time spent on inference. 
The average latency inference can be obtained by averaging the total time. 
For memory footprint, the flash usage is the sum of the value of \textit{Code}, \textit{RO-Data}, and \textit{RW-Data} reported by Keil when building and loading the submitted program on board.
With the collected practical performances, the final score of each submission can be calculated by Eqn.~\ref{func-fb}-~\ref{func-final-score}.

\section{TinyML Designs for VA Detection}
\label{sec_design}

In this section, the statistical analysis and discussion of the submitted AI/ML algorithms are presented, followed by the technical details of the top-3 teams' designs.

\subsection{Statistical Analysis of Submitted Designs}

\begin{figure}[t]
\includegraphics[width = 0.47\textwidth]{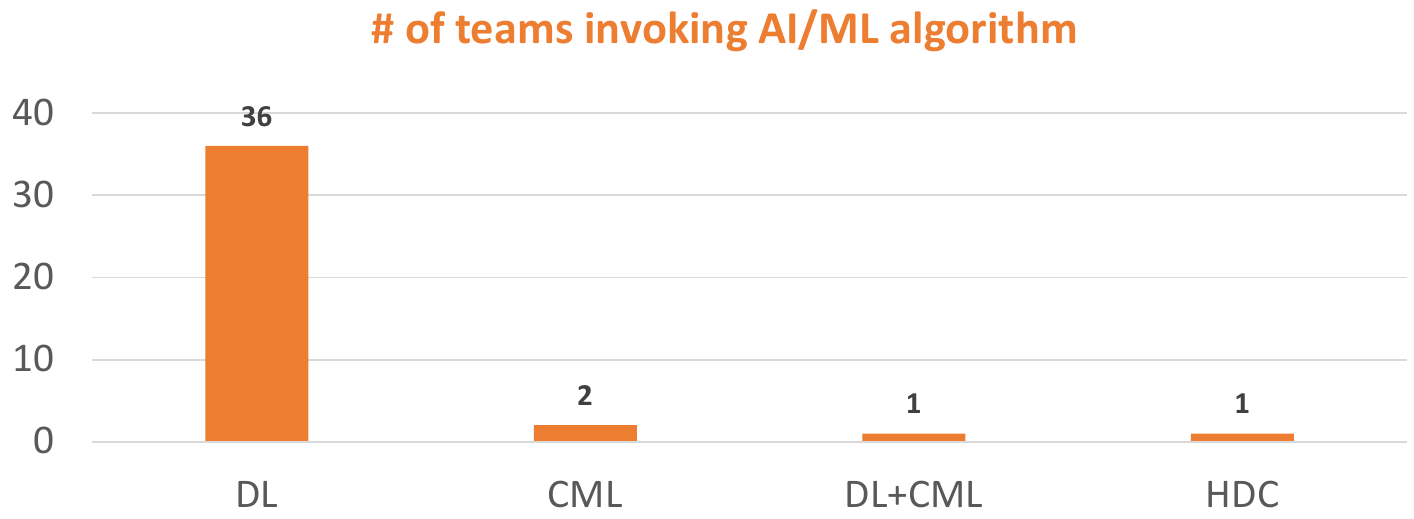}
\centering
\caption{Number of teams using different types of AI/ML algorithms. Note that DL represents Deep Learning, CML represents Conventional Machine Learning, and HDC represents Hyper-Dimensional Computing. 
}
\label{fig-ai}
\end{figure}

This paper reports the analysis of the top 40 teams' designs. 
Fig.~\ref{fig-ai} shows the distribution of the AI/ML algorithms applied to the team's design. 
As shown in the figure, deep learning (DL) dominates the choice of algorithms for the contest problem. 
All 36 out of the top 40 teams utilize Convolutional Neural Network (CNN) as the life-threatening VA detection model. 
As for conventional machine learning (CML), two teams choose Decision Tree to gain better practical on-board performances in terms of latency and memory footprint. 
In addition, one team applies DL and CML simultaneously in the design in order to gain a better overall performance. 
One team applies hyper-dimensional computing (HDC) to their design. 
The detailed results and analysis will be presented in Section~\ref{sec_results}.

\begin{figure}[t]
\centering
\subfigure[Optimizations for accuracy.]{
\includegraphics[width =0.47\textwidth]{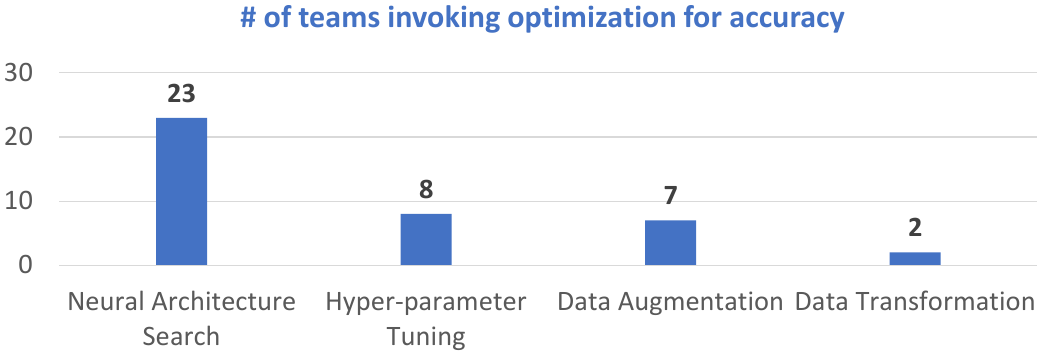} 
\label{fig-dis-tech-acc} \vspace{-10pt}}
\subfigure[Optimizations for latency and memory footprint.]{
\includegraphics[width =0.47\textwidth]{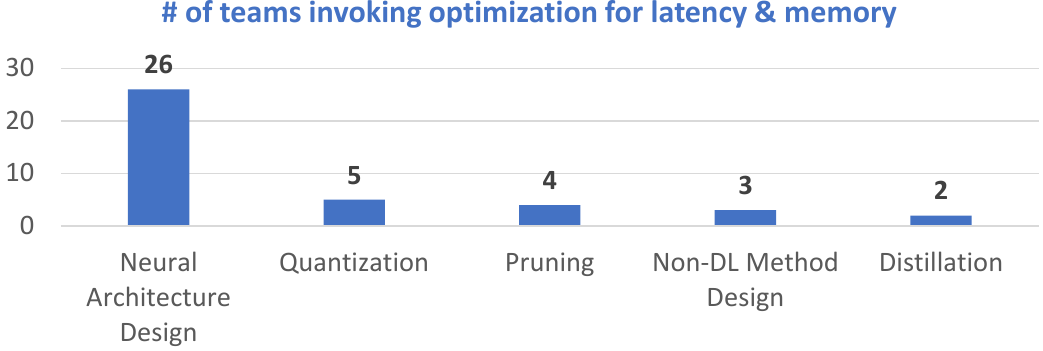}
\label{fig-dis-tech-latency}}
\caption{Distribution of the optimization technique applied for detection accuracy and practical performances.}
\label{fig-dis-tech}
\vspace{-10pt}
\end{figure}

Fig.~\ref{fig-dis-tech} shows the distributions of the specific optimization methods applied by the top 40 teams for better detection accuracy and practical performances respectively. 
As shown in Fig.~\ref{fig-dis-tech-acc}, from the perspective of model design, twenty-six teams apply neural architecture search to optimize the detection accuracy of the DL model. 
Neural architecture search here includes autoML, model selection based on evolution, empirical design, etc. 
There are eight teams choosing to fine-tune the hyper-parameters via Bayesian Search, grid search, and Ray package to gain a better performance. 
From the perspective of data, seven teams utilize data augmentation techniques such as signal flipping, resampling, and shifting to enrich the training dataset for better model generalization. 
There are two teams applying data transformation techniques, which transform the 1D signal into 2D frequency spectrum.

Fig.~\ref{fig-dis-tech-latency} shows the distribution of the teams applying different techniques for lower inference and memory footprint. 
Most teams (26 out of 40) utilize either autoML or empirical findings to design the CNN to fit the resource-constraint MCU platform. 
Five teams choose quantization while four teams choose pruning to reduce the model size and speed up the inference. 
There are two teams applying the model distillation technique to obtain the student model with a lower size while maintaining the detection accuracy. 
Moreover, three teams choose to implement the decision tree or HDC model on the board to reduce the computational complexity for better practical performances.

\subsection{Design of the Top 3 Teams}

The top 3 teams adopt various types of AI/ML algorithms.

\begin{figure*}[t]
\centering

\subfigure[Design of the 1st place team.]{
\includegraphics[width =0.65\textwidth]{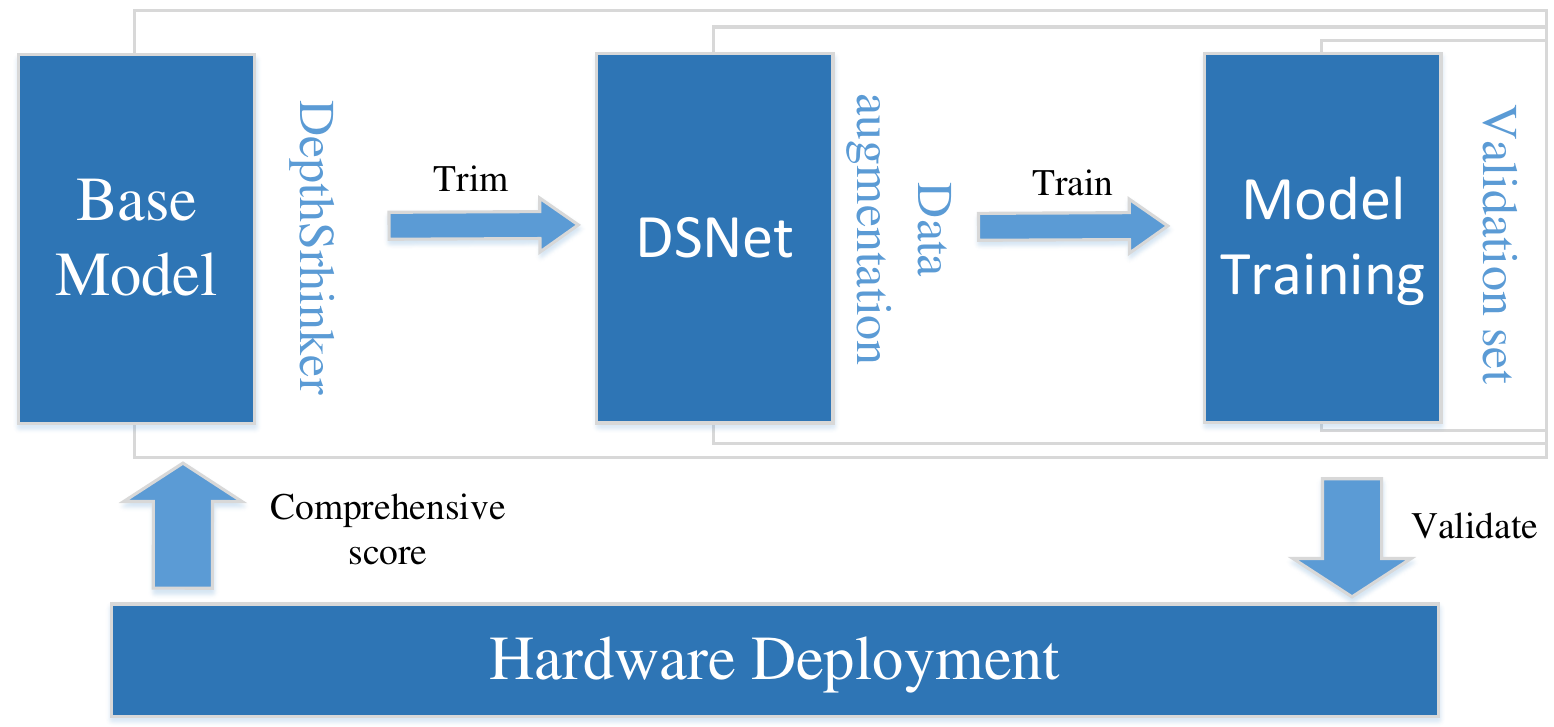} 
\label{fig-design-gatech} } 

\subfigure[Design of the 2nd place team.]{
\includegraphics[width =0.67\textwidth]{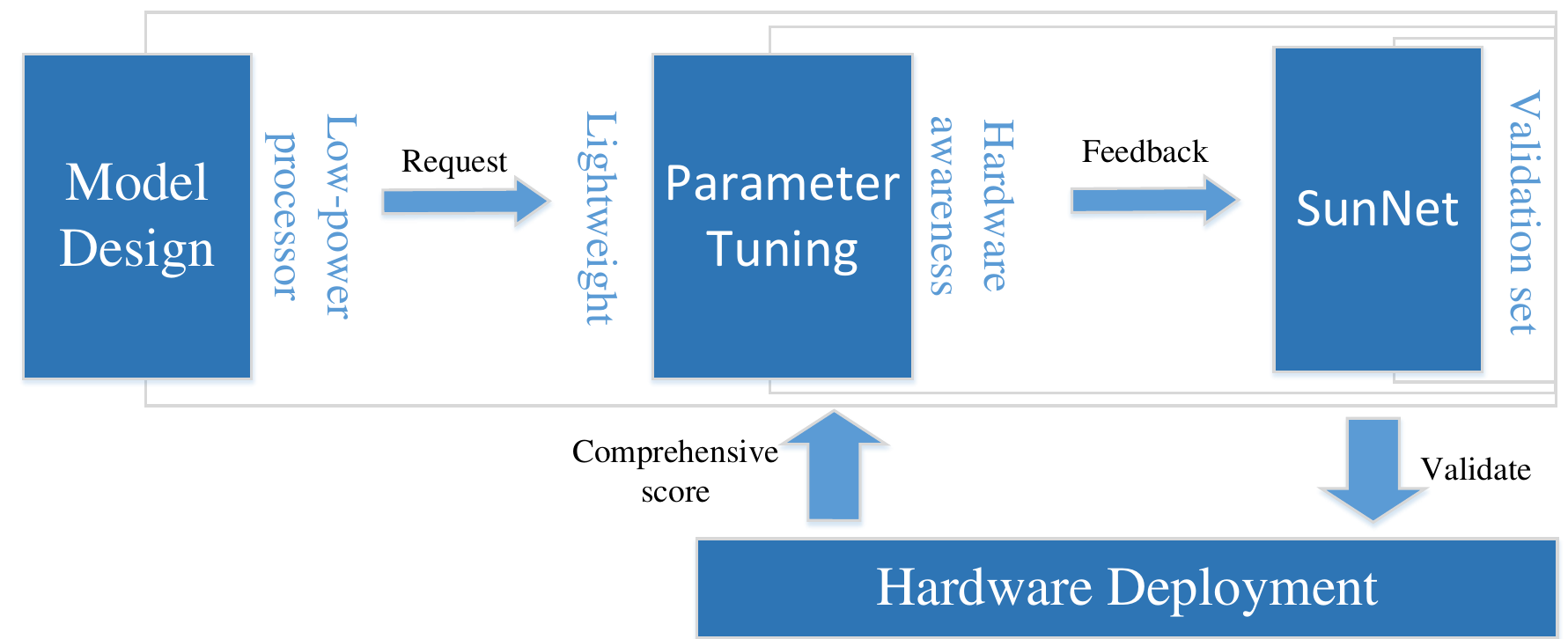}
\label{fig-design-seuer}}

\subfigure[Design of the 3rd place team.]{
\includegraphics[width =0.65\textwidth]{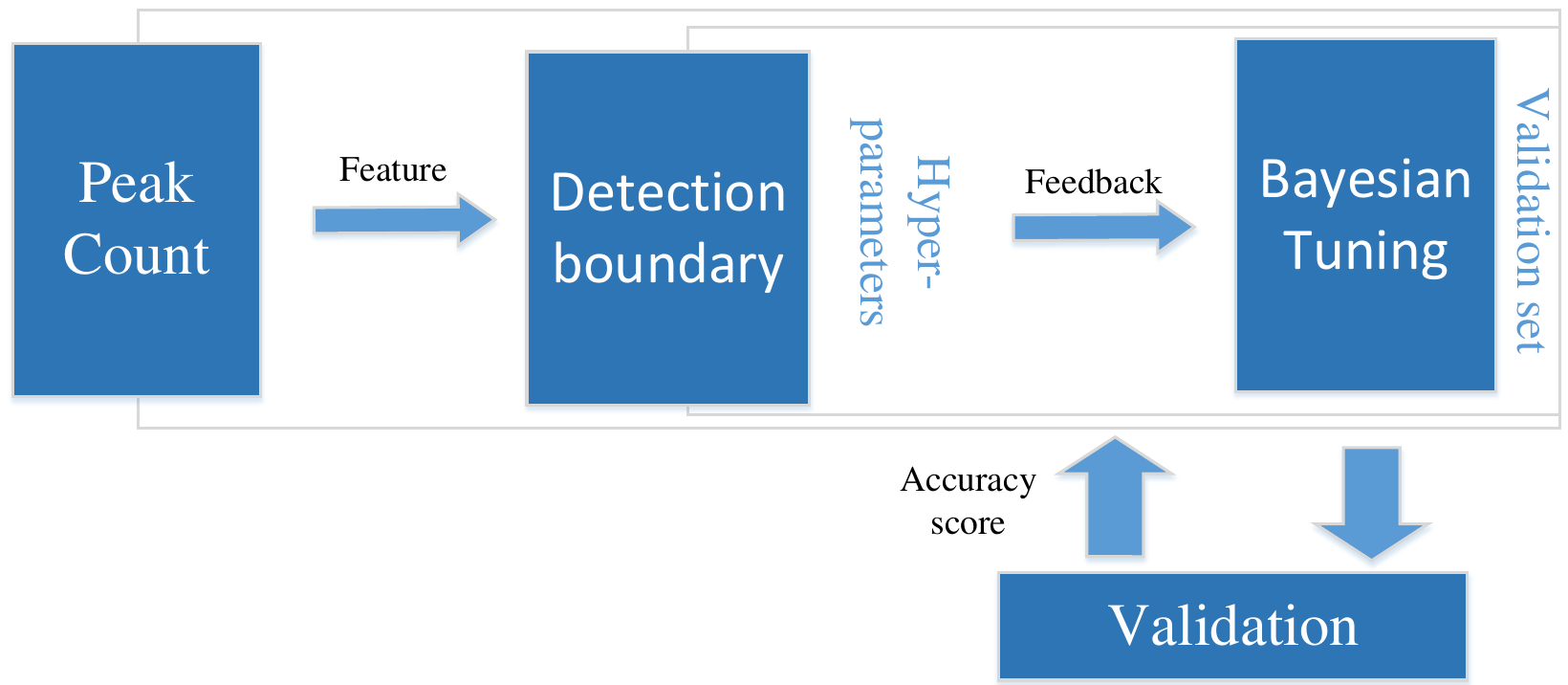}
\label{fig-design-mit}}

\vspace{-5pt}
\caption{The implementations of the top 3 teams ranked by the final score.}
\label{fig-design}
\vspace{-15pt}
\end{figure*}

\subsubsection{1st Place Team}
The first-place team utilizes a novel model compression paradigm to trim the model for developing hardware-efficient CNNs and applies several data augmentation techniques to enrich the training dataset for accurate VA detection. 
The design process is illustrated in Fig.~\ref{fig-design-gatech}.

Specifically, the team first proposes DSNet, a shallow CNN structure with large sparse kernels.
As shown in Fig.~\ref{fig-design-gatech}, inspired by DepthShrinker~\cite{fu2022depthshrinker}, the team obtain the DSNet structure by removing the basic building blocks of the baseline model (with five 1D-convolution layers and two fully-connected layers) provided by the contest organizers.
The approach transforms irregular computation patterns into dense ones with much-improved hardware utilization and thus improves execution efficiency. 
The trimmed network DSNet consists of one 1D-convolution layer (with kernel size = 85, stride = 32, channel = 3) and an MLP with hidden layers as [20, 10].

To improve the generalization of the trained model, as shown in Fig.~\ref{fig-design-gatech}, the 1st place team further applies data augmentation to the input IEGM segments, including flipping the signal and adding Gaussian noise.
With an augmented dataset, the team trains the model with 100 epochs, a cosine learning rate scheduler (with an initial learning rate of 0.0002, maximal learning rate of 0.0004, minimal learning of 0.0002, step of 100), an Adam optimizer, batch size of 32, and Stochastic Weight Averaging (SWA)~\cite{izmailov2018averaging} enabled after 10 epochs. 
More implementation details can be accessed at \url{https://github.com/GATECH-EIC/TinyML-Contest-Solution/tree/master}.

\subsubsection{2nd Place Team}
The second-place team adopts an iterative feedback loop based design for model development and deployment as shown in Fig.~\ref{fig-design-seuer}. 
The core idea of the design process is based on the observation that the network model and operators are required to be lightweight and easy to implement on the low-power processor.

Following the idea, the team proposes SunNet, a CNN architecture specifically designed for the ultra-low-power processor NUCLEO-L432KC. 
The network architecture searching process is based on hardware awareness by considering inference latency and network memory utilization, which are also the criteria in the final evaluation. 
As shown by Fig.~\ref{fig-design-seuer}, the team deploys the network on board with the help of STM32CubeMX and X-CUBE-AI toolchain. 
The practical performances (i.e., inference latency and memory footprint) and the detection performance (i.e., VA detection accuracy) are then utilized as the reference to enable the team to manually tune the hyper-parameters and model architecture to optimize the network.
The design procedure is conducted iteratively until the final comprehensive performance score of SunNet meets the criteria. 
The final SunNet consists of three 1D-convolution layers (with kernel size = 10, 9, 8, stride = 6, 5, 4, channel = 2, 4, 8) and an MLP with hidden layers as [32, 16].

With the searched network architecture, the team trains the model with 30 epochs, a cosine learning rate scheduler (with an initial learning rate of 0.01, a maximal learning rate of 0.01, minimal learning rate of 0.0001, step of 30), an Adam optimizer, batch size of 128. 
More implementation details can be accessed at \url{https://github.com/AiArtisan/ICCAD-TinyML-2nd-Place/tree/master}.

\begin{figure*}[t]
\includegraphics[width = 0.95\textwidth]{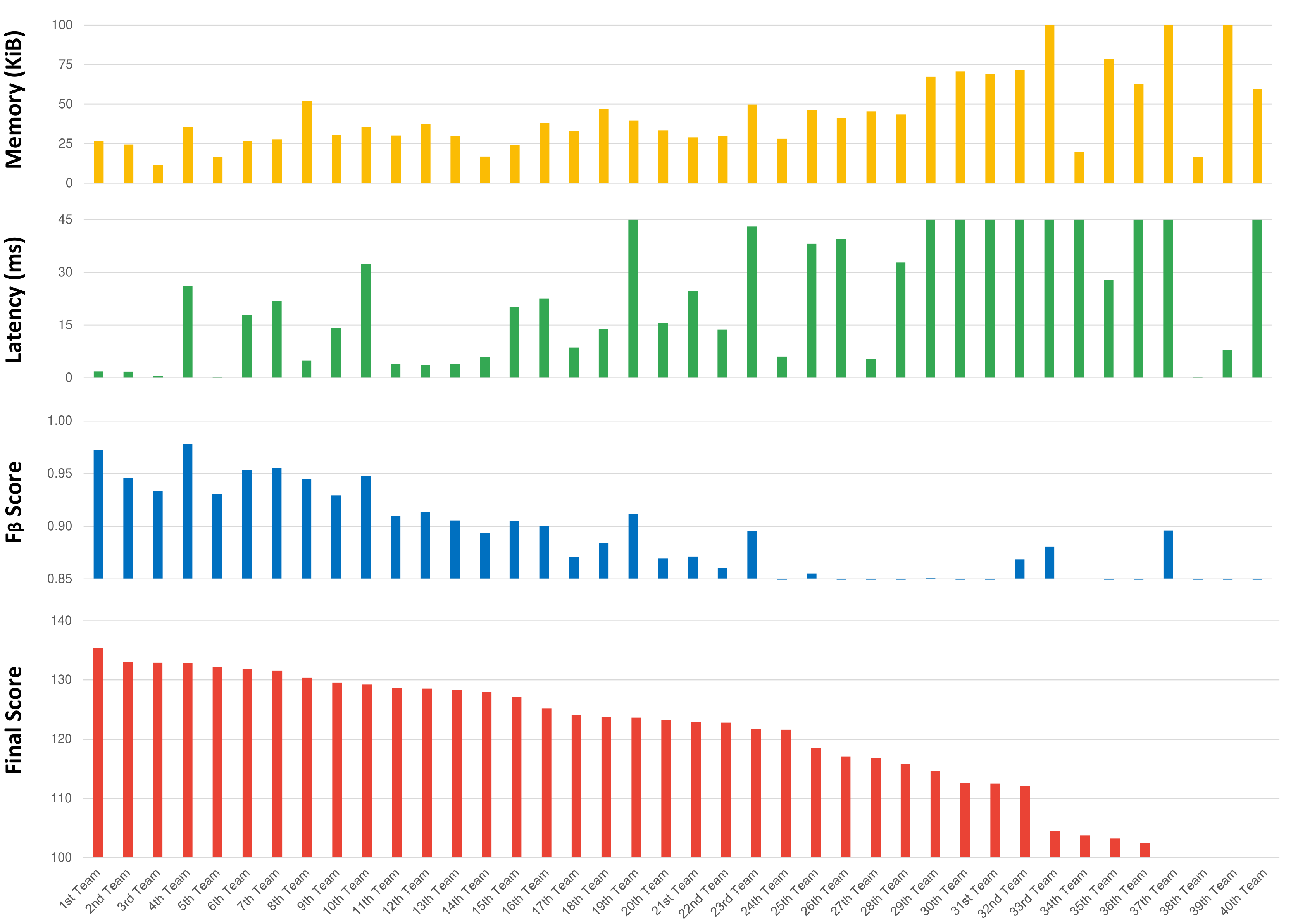}
\centering
\caption{Overall comprehensive performances of participating teams. The teams are ranked from high to low by the final score. The detailed scores and performances can be accessed at \url{https://tinymlcontest.github.io/TinyML-Design-Contest/Winners.html}. 
}
\vspace{-5pt}
\label{fig-overall-result}
\end{figure*}

\subsubsection{3rd Place Team}
The third-place team adopts the conventional machine learning approach (i,e., decision tree) to conduct VA detection. 
The objective of the team's design is to make a trade-off between the practical performances (i.e., latency and memory footprint) and the detection performance (i.e., VA detection accuracy). 
Utilizing conventional machine learning on hand-crafted features, the team chooses to gain a higher score in latency and memory footprint by scarifying the score in accuracy.

Specifically, as shown in Fig.~\ref{fig-design-mit}, the team first determines the number of peaks in the waveform (IEGM segment) as the feature.  
To precisely count the number of peaks, the team defines a factor $peak\_detection\_value$ with the value of $waveform.std() \times 2.0$. 
The factor scales the standard deviation of the waveform to obtain the peak detection value.
Points in the IEGM segment that are higher than this hurdle are recognized as peaks. 
The average number of peaks for different labels is shown in the first step in Fig.~\ref{fig-design-mit}.

Next, as shown in Fig.~\ref{fig-design-mit}, the team defines a threshold parameter as a decision boundary used to classify whether or not the IEGM segment is labeled with VAs (positive) or non-VAs (negative) based on the number of peaks.
In order to maximize the $F_{\beta}$ score, the team tunes the factor and threshold parameters of the decision tree via Bayesian search provided by Weights \& Biases Sweeps.
The final threshold is set to 9.215 and the factor is set to 2.095 for the highest comprehensive score. 
The implementation details of the third-place team can be accessed at \url{https://github.com/mit-han-lab/iccad-tinyml-open}.

\section{Results and Analysis}
\label{sec_results}

In this section, we will discuss the results in terms of various performance metrics. 
The final scores achieved by participating teams are presented and analyzed. 
Followed by the report of comprehensive performances, the effect of different AI/ML algorithms and frameworks are discussed with respect to detection accuracy, inference latency, and memory footprint. 

\begin{figure}[t]
\includegraphics[width = 0.47\textwidth]{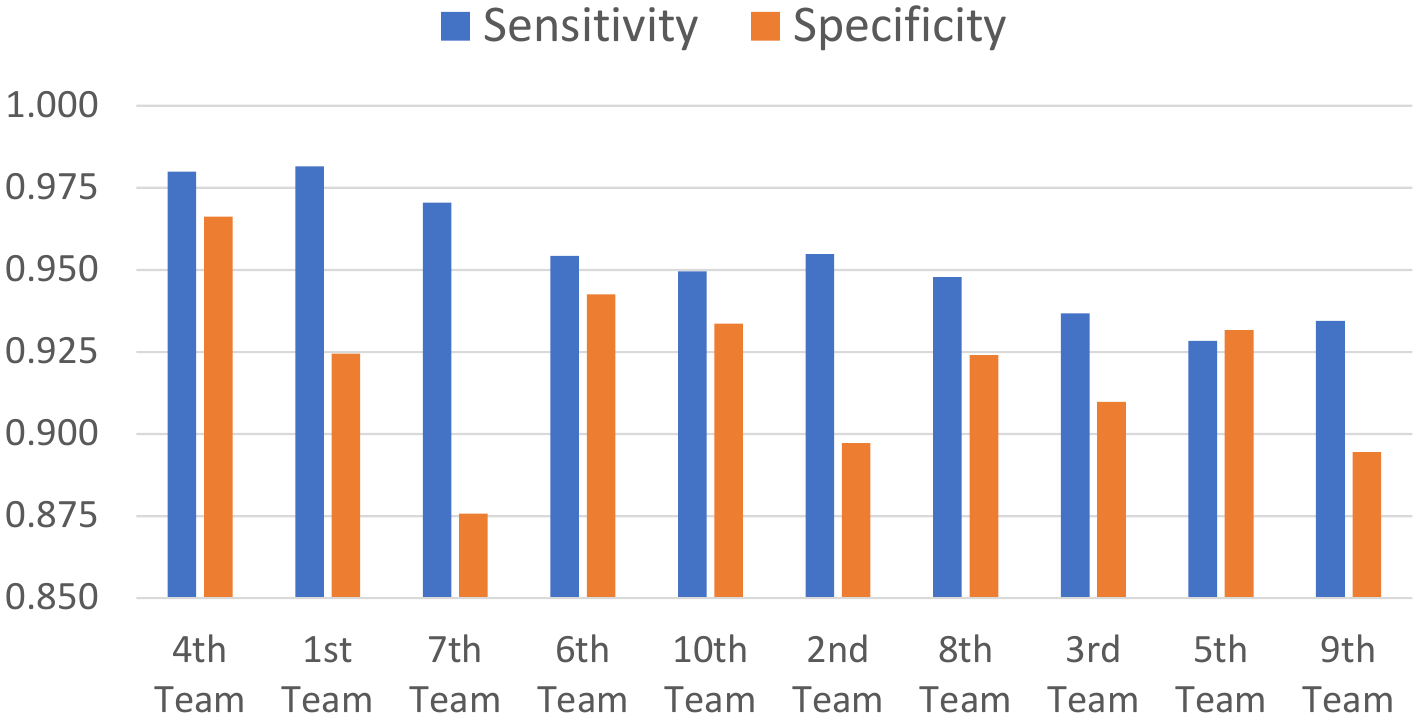}
\centering
\caption{The sensitivity and specificity of the VA classification task. Note that the teams are ranked with respect to the $F_{\beta}$ score and the leftmost one is the team with the highest $F_{\beta}$ score. 
The anonymous team's names are retrieved from Fig.~\ref{fig-overall-result}.
}
\vspace{-10pt}
\label{fig-result-spesen}
\end{figure}

\subsection{Overall Comprehensive Performances}

\begin{figure*}[t]
\includegraphics[width = 0.99\textwidth]{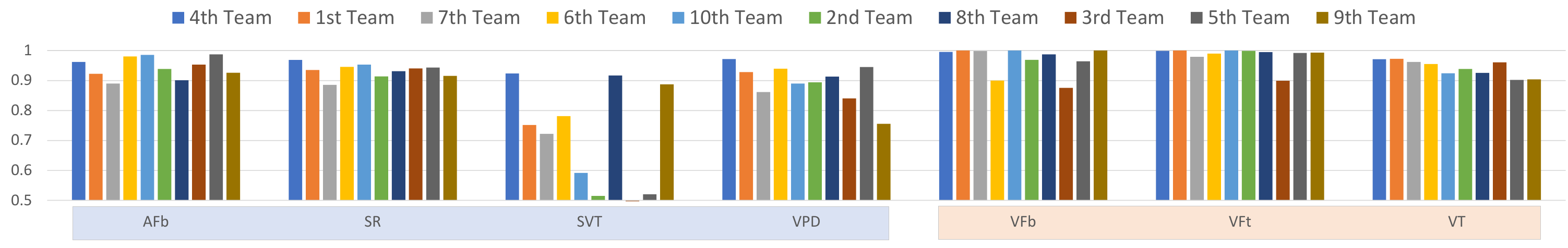}
\centering
\caption{The detection accuracy of each sub-category achieved by the top 10 teams with respect to the $F_{\beta}$ score. Note that the teams are ranked with respect to the $F_{\beta}$ score and the leftmost one is the team with the highest $F_{\beta}$ score. 
The anonymous team's names are retrieved from Fig.~\ref{fig-overall-result}.
}
\vspace{-5pt}
\label{fig-result-eachteam}
\end{figure*}

Fig.~\ref{fig-overall-result} shows the comprehensive performances of all participating teams ranked by the final score. 
As shown in the figure, the top 8 teams, which achieve a final score larger than 130, present a decent performance in terms of all three metrics. 
Their implementations are open-sourced and can be accessed at~\cite{1st-team, 2nd-team, 3rd-team, 4th-team, 5th-team, 6th-team, 7th-team, 8th-team}. 
Among these three metrics, detection accuracy (i.e., $F_{\beta}$) contributes more to the final score than the rest two metrics since the Eqn.~\ref{func-final-score} puts more weight (i.e., 100 vs 20) on the accuracy metric. 
As a result, all of the top 8 teams achieve the $F_{\beta}$ above 0.93 while only one team accomplishes it from the rest 9th-40th teams. 
The $F_{\beta}$ score reported in Fig.~\ref{fig-overall-result} also demonstrates a performance gap between the 10th and 11th teams, where the $F_{\beta}$ of the teams cannot exceed 0.91. 
It indicates that the high $F_{\beta}$ score is the base to achieve a better final score.
With a high $F_{\beta}$ score, low inference latency and memory footprint are the guarantees to obtain a high final score. 
As shown in Fig.~\ref{fig-overall-result}, the top 8 teams in the final score demonstrate decent performances in terms of inference latency and memory footprint.

To be more specific, the 1st place team achieves the final score of 135.43, which surpasses the 2nd place team by 2.5. 
The design of the 1st team presents outstanding comprehensive performances in terms of all metrics. 
It ranks the 2nd, 5th, and 8th place in $F_{\beta}$, latency, and memory footprint. 
In addition, the high $F_{\beta}$ (i.e., 0.972) of the team plays a key role in the final score ranking. 
The final score difference between consecutive teams ranking 2nd to 8th indicates fierce competition among the top teams. 
The 2nd place team chooses to balance all performance metrics, with 6th in $F_{\beta}$, 4th in latency, and 7th in memory footprint. 
Among the top teams, some of them choose to put more effort to optimize one or two specific metrics. 
For example, the 3rd place team and the 5th place team concentrate on practical performances in terms of latency and memory. 
Both teams achieve extremely low inference latency and memory footprint by sacrificing detection accuracy to some extent. 
The 4th place team, on the other hand, chooses to optimize the detection accuracy and achieves the highest $F_{\beta}$ to compensate for the other two metrics.

As shown in Fig.~\ref{fig-overall-result}, the setting of the final score in Eqn.~\ref{func-final-score} successfully motivate the teams to optimize the design toward better comprehensive performances in terms of detection accuracy, inference latency, and memory footprint.  
As for scalability, the models and approaches proposed by the participating teams can be applied to the biosignal data in other types. 
The model structure can be accordingly adjusted to fit the data dimension. 
The approaches such as data augmentation and neural architecture search are naturally generalized for different types of biosignal data.

\subsection{Performance by Detection Accuracy}

In addition to $F_{\beta}$, we further demonstrate the detailed detection performances in terms of various metrics reported by the participating teams. 
The discussion and analysis of the effect of different types of AI/ML algorithms are presented.

Fig.~\ref{fig-result-spesen} shows the sensitivity and specificity of the top 10 teams ranked by the $F_{\beta}$ score. 
Note that the condition positive is VAs and the condition negative is Non-VAs. 
As shown in the figure, the top 7 teams apply deep learning algorithms in their design. 
For example, the 4th place team achieves near-perfect sensitivity and specificity with the highest $F_{\beta}$ of 0.978. 
This team applies data augmentation to enrich the scale and representations of the training dataset and improve the generalization of the CNN model. 
The 1st place team chooses to conduct the neural architecture search to find the best-fit CNN model and data augmentation for better detection performances. 
As for the 3rd place team and the 5th place team, they both adopt conventional machine learning (i.e., decision tree) in their design.
Both teams experience a performance gap in sensitivity when compared with the 4th place team or the 1st place team. 
Moreover, as shown in Fig~.\ref{fig-result-spesen}, almost of teams achieve a better sensitivity over specificity. 
It indicates that the setting of $F_{\beta}$ defined in Eqn.~\ref{func-fb} successfully enforces the teams to put more weight on VA detection in order to gain a higher final score.

To further demonstrate the superiority of deep learning in detection performances, we further present the detection accuracy of each sub-category achieved by the top 10 teams in Fig.~\ref{fig-result-eachteam}. 
Note that the detection accuracy of the sub-category AFt is excluded since the number of AFt segments is less than 50, which cannot reach statistical significance. 
As shown in Fig.~\ref{fig-result-eachteam}, in the category of Non-VAs, deep learning based designs achieve relatively stable performances in terms of detection accuracy of each sub-category. 
As for the conventional machine learning based designs, the 3rd place team and the 5th place team could only reach around 50\% detection accuracy for SVT since both teams' designs highly rely on the feature of heart rate. 
However, the heart rate feature extracted from SVT and VT segments cannot effectively discriminate both arrhythmias.  
On the other hand, the teams adopting deep learning show a much better detection accuracy on SVT (i.e., 92.31\% by the 4th place team and 91.72\% by the 8th place team). 
In the category of VAs, teams achieve relatively similar performances over each sub-category. 
The top 3 teams demonstrate near-perfect detection accuracy on VFb, VFt, and VT.

Fig.~\ref{fig-result-spesen} and Fig.~\ref{fig-result-eachteam} both demonstrate the great potential of deep learning in life-threatening VA detection. 
In practice, inappropriate shock for misclassified SVT is also considered an unavoidable case since the principle of ICD is to avoid missing any life-threatening VAs. 
Reducing the misclassification rate on SVT while maintaining the detection accuracy on VAs is the goal for all ICD manufacturers. 
Deep learning helps to reduce the inappropriate shock rate since the CNN model could learn to extract the essential features which may not be obvious or visible in conventional detection methods of ICDs.

\subsection{Performance by Inference Latency}

\begin{figure}[t]
\vspace{-3pt}
\includegraphics[width = 0.47\textwidth]{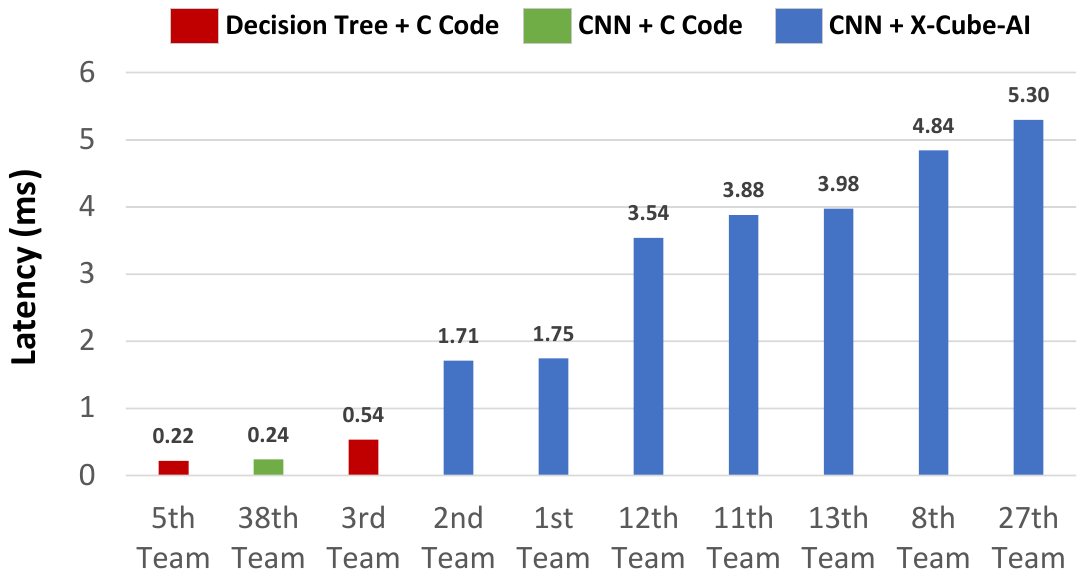}
\centering
\caption{The average inference latency of the top 10 teams. Note that the teams are ranked with respect to the latency and the leftmost one is the team with the lowest latency. 
The anonymous team's names are retrieved from Fig.~\ref{fig-overall-result}.
}
\vspace{-10pt}
\label{fig-result-latency}
\end{figure}

We further present the inference latency of selected teams and analyze the effect of different types of AI/ML algorithms and deployment tools on the latency.

Fig~.\ref{fig-result-latency} shows the average inference latency of the top 10 teams which run with the lowest inference latency. 
As shown in the figure,the 3rd place team and the 5th place team achieve 0.22 \textit{ms} and 0.54 \textit{ms}, which rank 1st and 3rd place in this track. 
Both teams adopt the decision tree as the detection model and implement the model with pure C code on board. 
Such design and implementation significantly reduce the computational overhead.
As for the 38th place team, it adopts CNN with an extremely simple structure and implements the model with pure C code on board to reduce the computational complexity. 
Note that the average inference latency achieved by these three teams is less than the lower bound (i.e. 1 \textit{ms}) set in Eqn.~\ref{func-latency}. 
Our setting rewards the teams with the inference latency lower than the lower bound to gain a score larger than 20 from the latency term. 
On the other hand, the marginal effects on the latency term further suppress the amount of the rewards in order to direct the teams to focus more on the optimization for detection accuracy.

As shown in Fig.~\ref{fig-result-latency}, nearly all of the teams except the 3rd place and the 5th place team adopt CNN in their design and deploy the model with the help of X-Cube-AI on board. 
In this way, the effective approach to reducing the inference latency is to optimize the CNN model itself. 
These teams apply quantization, pruning, and resource-aware neural architecture search to reduce computational complexity.

\subsection{Performance by Memory Footprint}

\begin{figure}[t]
\includegraphics[width = 0.47\textwidth]{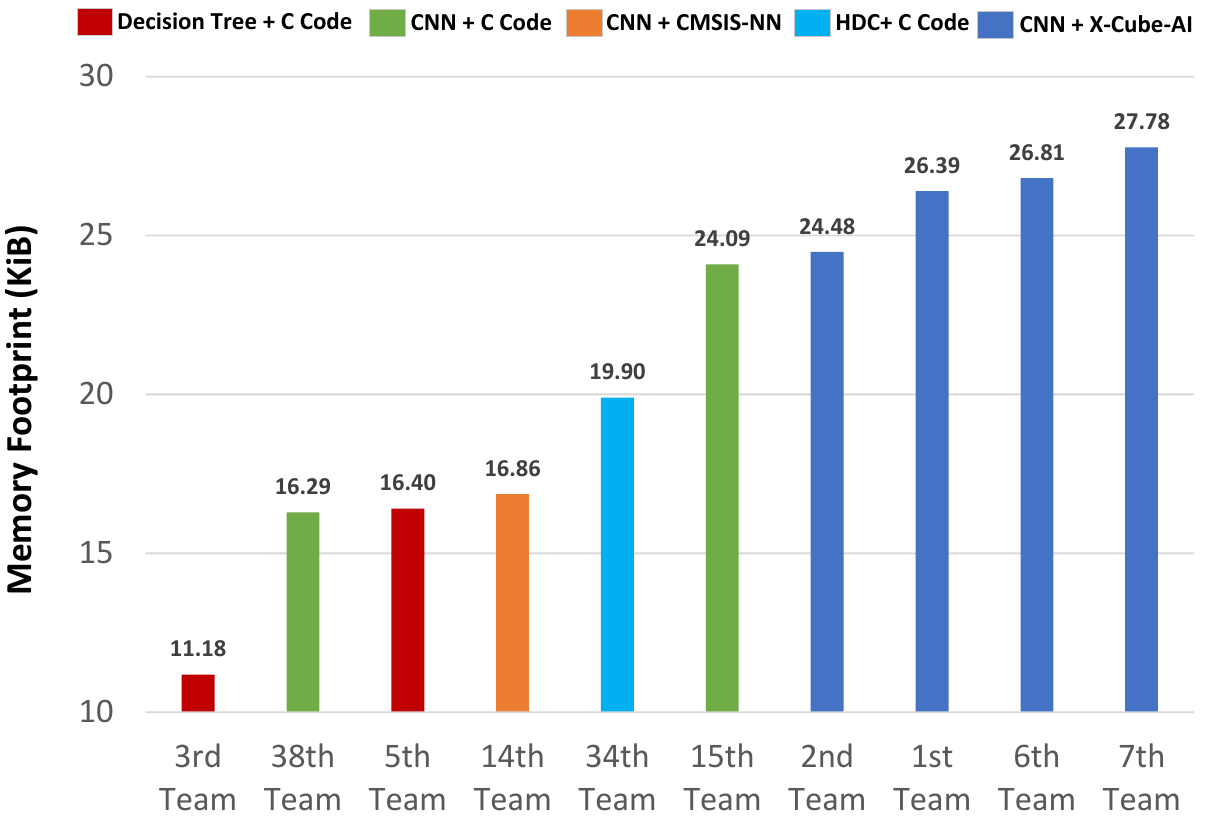}
\centering
\caption{The memory footprint of different teams. Note that the teams are ranked with respect to the flash occupation and the leftmost one is the team with the lowest flash occupation. 
}
\vspace{-10pt}
\label{fig-result-flash}
\end{figure}

We present the memory footprint of selected teams and analyze the effect of different types of AI/ML algorithms and deployment tools on the memory footprint. 
Fig~.\ref{fig-result-flash} shows the memory footprint of the top 10 teams which occupied the lowest flash memory. 
As shown in the figure, the 3rd place team uses only 11.18 \textit{KiB} space on board to execute their design. 
The 5th place team uses 16.40 \textit{KiB}, which ranks 3rd place. 
The adopted decision tree significantly reduces the memory overhead.
The 14th place team adopts quantization on the CNN model to shrink the model size and deploy it with CMSIS-NN on board. 
As shown in Fig.~\ref{fig-result-flash}, the rest of the teams adopt CNN and deploy the model with X-Cube-AI. 
These teams apply quantization, pruning, and resource-aware neural architecture search to shrink the model size. 

\subsection{Discussion}

\subsubsection{Advancement of TinyML in Healthcare}

The future vision of TinyML in healthcare is to enable the widespread deployment of AI/ML models on resource-constrained medical devices, such as wearable devices, implantable devices, and point-of-care devices~\cite{jia2023importance}. 
TinyML models can be used to monitor vital signs and health indicators, such as heart rate, blood pressure, and blood glucose levels, in real-time, and to alert patients and healthcare providers of any potential health issues.
TinyML models can be used to provide personalized therapy and care based on individual patient needs and health conditions.
This will allow for the creation of new and innovative solutions for healthcare problems, and the provision of personalized and continuous care to patients with significantly less expertise involved.
The future vision of TinyML in healthcare is to revolutionize the way medical devices and healthcare systems are designed, and to enable new and innovative solutions that can improve quality of life.

The way of designing an efficient TinyML system in the healthcare field shares a lot of commonalities with designing on the generic TinyML platforms. 
Hardware-aware neural architecture search~\cite{jiang2020standing, yang2020co}, model compression~\cite{wu2022fairprune}, work-in memory optimizations~\cite{wang2021lightweight}, and other approaches that are widely adopted for the generic TinyML platforms are also utilized to design the TinyML system for healthcare. 
On the other hand, absolute precision for detection to save lives or better life quality makes designing TinyML system for healthcare more challenging than the generic ones. 
In most healthcare applications, accuracy cannot be sacrificed for better practical performances while the hardware constraints of the mobile and implantable medical devices are even harsher than these of the generic TinyML platforms. 
For example, the lifetime requirement of ICD is typically 5-8 years while the detection method should run with 10-100 $\mu$A with extremely limited battery capacity. 
The challenging requirements of these medical devices greatly hinder the development of AI/ML for better healthcare.

TDC'22 provides an opportunity for teams to showcase their research and development in the field of TinyML in healthcare, and to demonstrate the latest advancements in this field.
Teams can explore new and innovative approaches to detecting and classifying different types of ventricular arrhythmias, and to improving the accuracy of these models.
The contest enables teams to learn best practices and techniques for developing tiny and efficient AI/ML models that can run on resource-constrained devices.

\subsubsection{Future Direction for TinyML in Healthcare}
TDC'22 further reveals several essential perspectives on the future direction of TinyML in healthcare. 

Deep learning and conventional machine learning are not contradictory options for TinyML design in healthcare. 
As shown by the performances in terms of detection accuracy, latency, and memory footprint, deep learning based methods can learn to extract essential features that are not visible or noticeable in conventional feature engineering. 
These extracted features could further improve the detection performances. 
On the other hand, conventional machine learning based methods benefit from the low computational complexity. 
These methods are relatively simple to be implemented and deployed on board, with lower memory footprint and execution latency. 
Therefore, it is natural and practical to combine these two types of AI/ML algorithms to gain better comprehensive performances.

The resource constraints on tiny and embedded devices are the critical obstacle for larger model deployment for better accuracy. 
The future solutions can be concluded as follows: 
1) New toolchain for model deployment. The toolchain should be able to provide full-stack solutions, including model structure design, model compression, and deployment with low-level optimization; 
2) Tiny and embedded platform with emerging devices to enrich hardware resources~\cite{yan2022swim}. The platform equipped with more resources could further improve detection performances;  
3) Cross-layer co-design for the specific health application or even individual. Co-optimize the components ranging from devices, circuits, architecture, system, and algorithm design instead of developing each component independently and separately~\cite{jiang2020device, jiang2020hardware};  
4) Self-adaptation on both hardware and software levels. Adjust the hardware and software settings to make the medical device tailored to each individual's changing circumstances for better accuracy and efficiency.

In the scenarios of health applications, non-iid health data is universal for model training and practical testing such as the patient-wise data partition applied in TDC'22. 
The new optimizations targeting model generalization such as meta-learning are the key solution to address the challenge~\cite{jia2022personalized}. 
In addition, the metric that objectively measures the generalization of the model is essential for TinyML in healthcare.

\section{Conclusion}
\label{sec_conclusion}

In this paper, we present the 1st ACM/IEEE TinyML Design Contest for life-threatening ventricular arrhythmia detection on ICDs. Specifically, we introduce the contest problem with medical background and the significance of applying TinyML in the application. 
We further describe the IEGM dataset, evaluation metrics for comprehensive performances, and the setup for on-board evaluation. 
We also present the design of participating teams as well as the analysis of the adopted AI/ML algorithms. 
The analysis of the results achieved by participating teams is presented together with an in-depth discussion for researchers and engineers to further contribute to TinyML in health.

\section*{Acknowledgement}

TDC'22 is sponsored by IEEE Council on Electronic Design Automation (CEDA) and ACM Special Interest Group on Design Automation (SIGDA).

\bibliographystyle{IEEEtran} 
\bibliography{./bib/ref.bib}

\vskip -20pt plus -1fil 

\begin{IEEEbiography}[{\includegraphics[width=1in,height=1.35in]{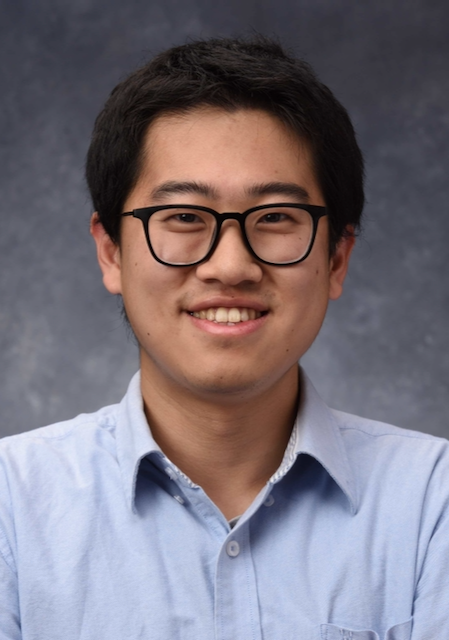}}]{Zhenge Jia}
is currently a Postdoc in the Department of Computer Science and Engineering at the University of Notre Dame. He obtained his Ph.D. degree in Electrical and Computer Engineering at the University of Pittsburgh in 2022. He received his B.S. degree in Engineering and Computer Science at Australian National University in 2017. His research interests include personalized deep learning and on-device deep learning in healthcare. He published more than 10 papers in Nature Machine Intelligence, DAC, ICCAD, TCAD, IJCAI and received the Second Place Award in Ph.D. Forum at DAC 2023. He has served as TPC member in ICCAD and reviewer for TCAD, TNNLS, Nature Scientific Reports, JETC, TCPS, ESL, TCAS-II, etc. 
\end{IEEEbiography}

\vskip -10pt plus -1fil 

\begin{IEEEbiography}[{\includegraphics[width=1in,height=1.2in]{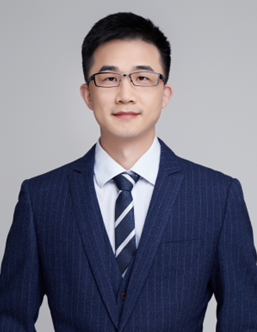}}]{Dawei Li}
is currently an assistant professor at South Central Minzu University. He received his B.S. degree from Wuhan University of
Science and Technology in 2010, and his Ph.D. degree from
Huazhong University of Science and Technology in 2017, both
in Electrical Engineering. His current research interests include
Low-power front-end, LDO, and VCOs. 
\end{IEEEbiography}
\vskip -10pt plus -1fil 

\begin{IEEEbiography}[{\includegraphics[width=1in,height=1.2in]{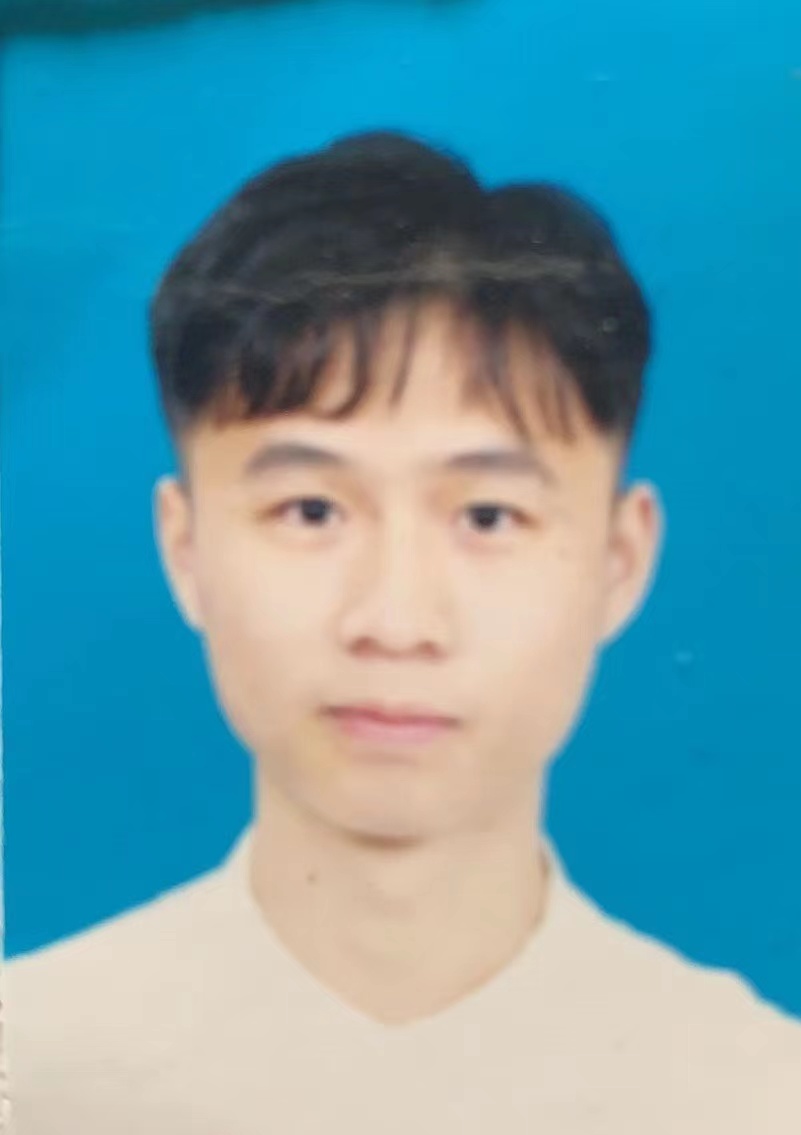}}]{Cong Liu}
is currently studying for a master's degree at South  Central Minzu University. He received his B.S. degree from South Central Minzu University in 2016. His current research interests include biomedical signal processing, computer vision, and privacy protection.
\end{IEEEbiography}
\vskip -10pt plus -1fil 

\begin{IEEEbiography}[{\includegraphics[width=1in,height=1.2in]{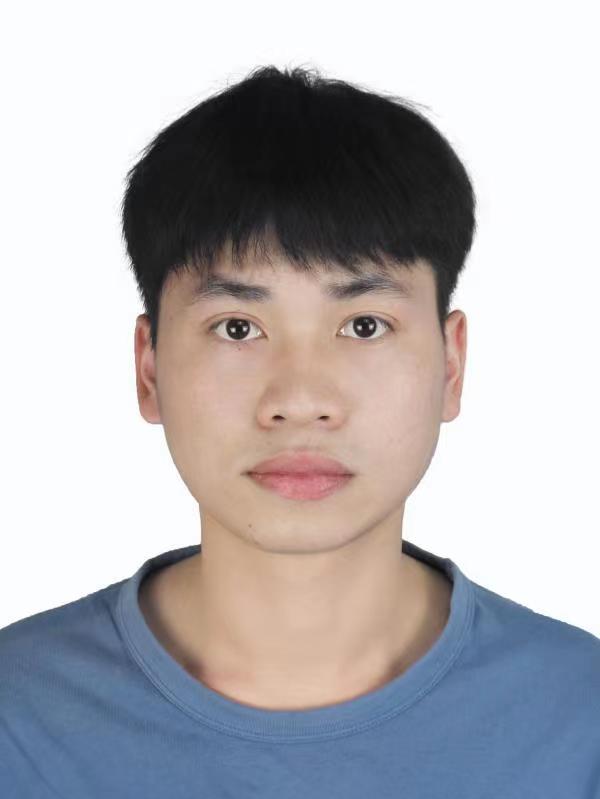}}]{Liqi Liao}
is currently studying for a master's degree at South central Minzu university. He received his B.S. degree from Inner Mongolia University for Nationalities in 2021. His current research interests include biomedical signal processing and computer vision.
\end{IEEEbiography}
\vskip -10pt plus -1fil

\begin{IEEEbiography}[{\includegraphics[width=1in,height=1.1in]{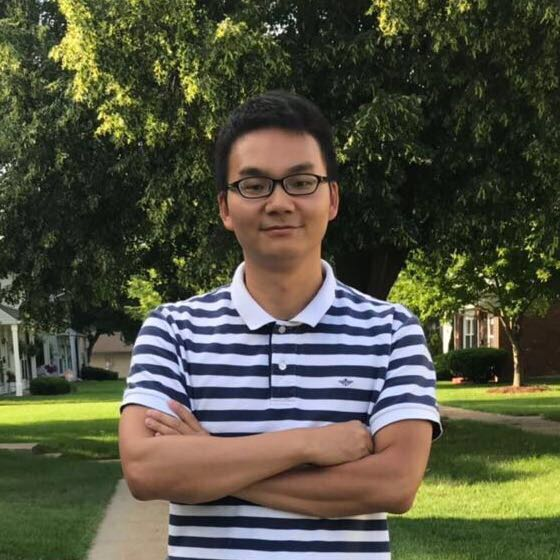}}]{Xiaowei Xu} received the BS and PhD degrees in
electronic science and technology from the Huazhong University of Science and Technology, Wuhan, China, in 2011 and 2016, respectively. He is currently an associate professor with Guangdong Cardiovascular Institute, Guangdong Provincial People’s Hospital, Guangzhou, China. He worked as a post-doc researcher with the University of Notre Dame, IN, USA from 2016 to 2019.
His research interests include deep learning, and
medical image segmentation. He was a recipient
of DAC system design contest special service recognition reward in 2018
and outstanding contribution in reviewing, Integration, the VLSI journal in
2017. He has served as TPC members in ICCD, ICCAD, ISVLSI and
ISQED.
\end{IEEEbiography}
\vskip -10pt plus -1fil 

\begin{IEEEbiography}[{\includegraphics[width=1in,height=1.2in]{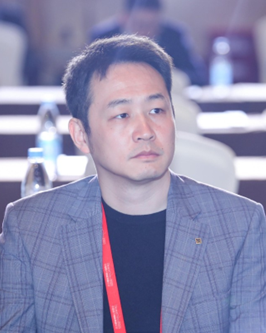}}]{Lichuan Ping}
is the founder and CEO of Singular Medical. He was Visit Research Assisant Professor at University of Notre Dame from 2018 to 2019. He received his B.S. degree in Electrical Engineering from Harbin Engineering University in 2006. He received his Ph.D. degree in Signal and Information Processing from University of Chinese Academy of Sciences in 2011.
\end{IEEEbiography}
\vskip -10pt plus -1fil 

\begin{IEEEbiography}[{\includegraphics[width=1in,height=1.2in]{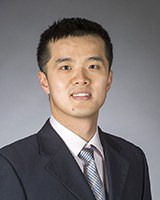}}]{Yiyu Shi}
is currently a professor in the Department of Computer Science and Engineering at the University of Notre Dame, the site director of National Science Foundation I/UCRC Alternative and Sustainable Intelligent Computing, and the director of the Sustainable Computing Lab (SCL). He received his B.S. in Electronic Engineering from Tsinghua University, Beijing, China in 2005, the M.S and Ph.D. degree in Electrical Engineering from the University of California, Los Angeles in 2007 and 2009 respectively. His current research interests focus on hardware intelligence and biomedical applications. In recognition of his research, more than a dozen of his papers have been nominated for or awarded as the best paper in top journals and conferences, including the 2021 IEEE Transactions on Computer-Aided Design Donald O Pederson Best Paper Award. He is also the recipient of Facebook Research Award, NSF CAREER Award, IEEE Region 5 Outstanding Individual Achievement Award, IEEE Computer Society Mid-Career Research Achievement Award, among others. He has served on the technical program committee of many international conferences. He is the deputy editor-in-chief of IEEE VLSI CAS Newsletter, and an associate editor of various IEEE and ACM journals. He is an IEEE CEDA distinguished lecturer and an ACM distinguished speaker.
\end{IEEEbiography}
\vspace{-15pt}

\end{document}